\documentclass[10pt,final,doublecolumn]{IEEEtran}
\usepackage{amsmath}
\usepackage{amssymb}
\usepackage{amsfonts}
\usepackage{latexsym}
\usepackage{graphicx}
\usepackage{bbding}
\usepackage{enumerate}
\usepackage{subeqnarray}
\usepackage{multicol}
\usepackage{color}
\usepackage{slashbox}
\usepackage{setspace}
\usepackage{mathrsfs}
\usepackage{array}
\usepackage{algorithm,algorithmic}
\usepackage{colortbl}
\usepackage{bm}
\IEEEoverridecommandlockouts
\allowdisplaybreaks[4]

\begin{document}
\title{Integrating Sensing, Computing, and Communication in 6G Wireless Networks: Design and Optimization}
\author{Qiao Qi, Xiaoming Chen, Ata Khalili, Caijun Zhong, Zhaoyang Zhang, and Derrick Wing Kwan Ng
\thanks{Qiao Qi, Xiaoming Chen, Caijun Zhong, and Zhaoyang Zhang are with the College of Information Science and Electronic Engineering, Zhejiang University, Hangzhou 310027, China. Ata Khalili is with the Institute for Digital Communications, Friedrich-Alexander-University Erlangen-Nurnberg, Erlangen 91054, Germany. Derrick Wing Kwan Ng is with the School of Electrical Engineering and Telecommunications, the University of New South Wales, NSW 2052, Australia.}}\maketitle

\begin{abstract}
The roll-out of various emerging wireless services has triggered the need for the sixth-generation (6G) wireless networks to provide functions of target sensing, intelligent computing and information communication over the same radio spectrum. In this paper, we provide a unified framework integrating sensing, computing, and communication to optimize limited system resource for 6G wireless networks. In particular, two typical joint beamforming design algorithms are derived based on multi-objective optimization problems (MOOP) with the goals of the weighted overall performance maximization and the total transmit power minimization, respectively. Extensive simulation results validate the effectiveness of the proposed algorithms. Moreover, the impacts of key system parameters are revealed to provide useful insights for the design of integrated sensing, computing, and communication (ISCC).
\end{abstract}


%introduction
\section{Introduction}
The demands of various emerging wireless services, e.g., unmanned vehicle, holographic communication, and extended reality, require wireless networks to provide integrated functions of target sensing, intelligent computing and information communication in a timely manner \cite{6G1}. In current fifth-generation (5G) wireless networks, these functions are separated and irreciprocal for simplicity. Generally speaking, 5G wireless networks can provide enhanced mobile broadband communications \cite{5G1}, cloud and edge computing \cite{5G2} but with limited environment sensing capability \cite{5G3}. As a result, these functions in 5G wireless networks cannot satisfy the requirements of emerging wireless services. In this context, it is desired to study the integrated sensing, computing, and communication (ISCC) to improve the utilization of limited system resources for the sixth-generation (6G) wireless networks \cite{6G2}.

In fact, integrated sensing and communication (ISAC) has received considerable attentions in both academic and industry \cite{ISAC1}-\cite{ISAC3}. At the preliminary research stage, as a special case of ISAC, spectrum sharing between radar sensing and wireless communication has been investigated to improve the spectral efficiency. In particular, the co-channel interference between radar and communication systems was carefully mitigated via coordination of the two separated systems \cite{ISAC4}. However, the performance gain of orthogonal spectrum sharing-based radar sensing and wireless communication is limited. As a remedy, radar-centric communication integration was proposed by conveying information over radar signals concurrently via the same spectrum. For example, the amplitude and the phase of radar signals were controlled modulating some information bits \cite{ISAC5}. Since radar signals can be changed to a limited extent, the information transmission rate of radar-centric communication integration is severely unsatisfactory. In contrary, sensing-aided communication was also proposed to enhance the performance of wireless communication by using the sensed information. For instance,  the authors in \cite{ISAC6} designed an effective beam selection scheme for millimeter-wave communications based on the sensed environment information, e.g., the shape, the position, and the materials of surrounding building/car/tree by using a camera. Yet, the functions of sensing in sensing-aided communication are constrained. In this context, ISAC based on the integrated wireless architecture is regarded as a promising approach to enable 6G wireless networks \cite{ISAC7}. By optimizing the wireless transceivers, it is possible to improve both the sensing and communication performance simultaneously. In particular, since the base station (BS) will be equipped with a large-scale antenna array for 6G wireless networks, the superposition coded sensing and communication signal can be effectively separated by using spatial beamforming \cite{ISAC8}. Inspired by this, the authors in \cite{ISAC9} studied a joint communication and sensing 6G cellular system and discussed the optimal waveform design for communication as well as sensing by exploiting ultra-high terahertz spectrum and ultra-massive antenna array.

Meanwhile, integrated computing and communication has attracted significant interests due to the demands of huge-volume data processing \cite{ICC1,ICC2}. Originally, cloud computing was commonly adopted by exploring the powerful computing capability of cloud servers. For example, the authors in \cite{ICC3} jointly optimized the precoder of communication signals and the computational resources of cloud servers to minimize the energy consumption at the terminals. Yet, with the increasing of data volume, cloud computing leads to prohibitive communication load and latency. To address this issue, edge computing is applied to wireless networks \cite{Edge1,Edge2}. By processing the data at the edge of wireless networks, e.g., the BS, it is likely to increase the efficiency of computing and communication. For instance, the transmit power of communication signals and computational resources of edge servers in a multi-cell wireless network were jointly optimized to maximize the weighted sum of reductions in time and cost during task completion in \cite{ICC4}. Recently, computing is implemented via cooperation among the BS and the targeted devices, namely over-the-air computation, to further increase the computing efficiency and to decrease the communication load \cite{ICC5}. Specifically, by taking advantage of the superposition nature of wireless multiple access channels, numerous computing functions can be realized at the BS by coordinating the transmit signals of the end devices. In \cite{ICC6}, the authors jointly optimized transmit and receive beamforming to combat the interference between computing and communication signals. As the processing capability of the end devices increases, over-the-air computation and federate learning are gradually combined to enable intelligent computing, namely over-the-air federate learning (AirFL) \cite{ICC7}. In practice, AirFL trains local computing models at the end devices with their raw data and aggregates a global computing model at the BS by exploiting the over-the-air computation. In \cite{ICC8}, the authors derived the tradeoff between communication and computing, and designed a broadband aggregation algorithm for analog AirFL to reduce the communication latency.

Moreover, integrated sensing and computing has also become the research focus with the widespread applications of sensing services. In \cite{ISC1}, a novel distributed framework was developed by combing edge computing and distributed deep learning for urban environment sensing, which can effectively improve the computing efficiency and the sensing accuracy. The authors in \cite{ISC2} developed a multiple-input multiple-output (MIMO) over-the-air computation technique for realizing multi-modal sensing with high-mobility. The equalization scheme based on channel feedback was designed to minimize the computation error. In addition, a real-time big data analytical framework for remote satellite sensing service was put forward to enhance the computation efficiency in the context of massive data \cite{ISC3}.

Commonly, previous works attempt to integrate two functions in an effective way. In order to support emerging intelligent applications and services in 6G wireless networks, it is necessary to integrate sensing, computing and communication simultaneously. In this paper, we aim to provide feasible and effective integration schemes by exploiting available resources of 6G wireless networks. The contributions of this paper are three-fold:

\begin{enumerate}

\item We put forward a general design framework of ISCC for 6G wireless networks. Massive MIMO techniques enable the networks to perform multiple-target sensing, multiple-dimension computing, and multiple-stream communication simultaneously over the same wireless resources.

\item We design two typical joint beamforming algorithms for ISCC from the perspectives of maximizing the overall performance subject to transmit power constraints and minimizing the total transmit power consumption subject to performance requirements, respectively.

\item We reveal the impacts of key system parameters on the overall performance of ISCC, and provide useful guideline for practical system design.
\end{enumerate}

The rest of this paper outlined as follows: Section II briefly introduces a 6G wireless network integrating sensing, computing and communication. Section III presents two joint beamforming design algorithms of the weighted overall performance maximization and the total transmit power minimization, respectively. Section IV provides extensive simulations to verify the effectiveness of the proposed algorithms and Section V concludes the paper.

\emph{Notations}: We use $\Omega_N=\{1,2,\ldots,N\}$ to denote a set for given natural number $N$, bold upper (lower) letters to denote matrices (column vectors), $\|\cdot\|$ to denote $L_2$-norm, $|\cdot|$ to denote the absolute value of a scalar or the size of a dataset, and $\mathbb{E}\{\cdot\}$ to denote expectation.  ${{\mathbb{R}}^{m\times n}}$ and ${{\mathbb{C}}^{m\times n}}$ represent the sets of $m$-by-$n$ dimensional real matrix and complex matrix, respectively.  $(\cdot)^H$, $\text{tr}(\cdot)$, $\text{Rank}(\cdot)$ indicates the conjugate transpose, the trace, and the rank of a matrix, respectively.

%system-model
\section{System Model}
\begin{figure}[!h]
 \centering
\includegraphics [width=0.5\textwidth] {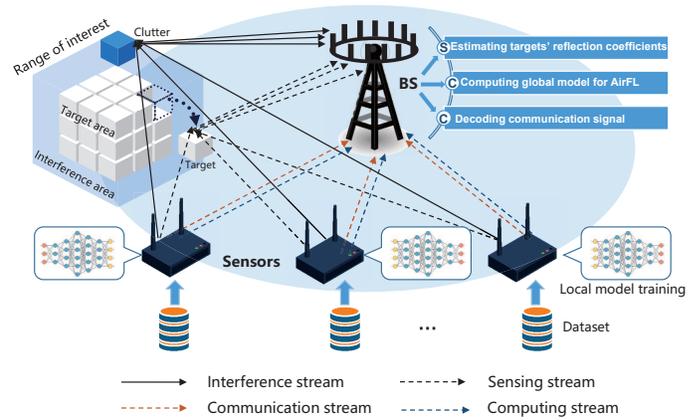}
\caption {System model for 6G wireless network integrating sensing, computing and communication.}
\label{Fig1}
\end{figure}
 We consider a 6G wireless network, c.f. Fig. \ref{Fig1}, consisting of a BS equipped with $N$ antennas, and $K$ multi-function sensors equipped with $M$ antennas each, where $KM\leq N$. The range of interest (ROI), including target sensing area and interference area, is partitioned into multiple cubes with equal size, each of which denotes a pixel point\footnote{Note that the size of the pixel represents the accuracy of sensing. In fact, pixels are not necessarily cubic, and different side lengths can be set individually to change the accuracy of different dimensions. According to the requirement of sensing accuracy, the size and the shape of the pixels can be determined based on the wavelength of the signal, the number of BS antennas, the overall size of the ROI and the wireless environment.} \cite{image}. Without loss of generality, there exist $O$ clutters\footnote{In general, the clutters can be considered as scatters over wireless channels. Hence, the number of clutters and channel state information can be acquired by channel estimation \cite{clutter}.} within interference range around the targets as interference for sensing, such a 6G wireless network has three fundamental tasks, namely target sensing, model computing, and information communication, which are collaboratively completed by multiple intelligent sensors and a multi-antenna BS. Specifically, based on the acquired environmental information\footnote{The environmental information can be roughly known by environmental detection. For example, channel conditions can be obtained by channel estimation \cite{environ1} and the location of the ROI can be obtained by beam sweeping \cite{environ2}.}, each intelligent sensor first performs spatial directional beamforming to control the transmit directions of the superposition of its sensing, computing, and communication signals, such that the signals reach the BS adaptively to the channel conditions. As for target sensing, the BS receives the reflected signals containing the environmental information of the target area and then designs the corresponding sensing receivers to estimate the reflection coefficients of targets\footnote{Since all the pixel points of the ROI have the same size, shape and alignment, the reflection coefficients are the only recognition features in this work. For example, the BS can perform target classification \cite{target} or target imaging \cite{imaging} by accurately obtaining their reflection coefficients.}. As for model computing, by exploiting the computing ability of intelligent sensors that train local models with their datasets, the BS designs the corresponding computing receivers to obtain the desired computing signals to facilitate the global model via AirFL. As for information communication, the BS designs the corresponding communication receivers to decode the communication signals of each sensor for data transmission. With the above framework, we can integrate sensing, computing and communication by a unified transceiver.

Without loss of generality, we assume that the $k$-th sensor constructs the sensing signals $\mathbf{s}_{k}^{\text{sens}}={{\left[ s_{k,1}^{\text{sens}},\ldots,s_{k,i}^{\text{sens}},\ldots ,s_{k,I}^{\text{sens}} \right]}^{T}}\in {\mathbb{R}^{I\times 1}}$ to sense $I$ targets, namely $I$ pixel points in the target area, generates local model parameters $\mathbf{s}_{k}^{\text{comp}}={{\left[ s_{k,1}^{\text{comp}},\ldots,s_{k,l}^{\text{comp}},\ldots ,s_{k,L}^{\text{comp}} \right]}^{T}}\in {\mathbb{R}^{L\times \text{1}}}$ as computing signals to perform AirFL, and logs data $\mathbf{s}_{k}^{\text{comm}}={{\left[ s_{k,1}^{\text{comm}},\ldots,s_{k,j}^{\text{comm}},\ldots ,s_{k,J}^{\text{comm}} \right]}^{T}}\in {\mathbb{R}^{J\times 1}}$ as communication signals to transfer the information in each time slot.
Note that the functions of theses three types of signals are different, where ${s}_{k,i}^{\text{sens}}$ is the sensing signal for the $i$-th target from the $k$-th sensor, which does not carry any data, but is just used to estimate the reflection coefficients of the targets, ${s}_{k,l}^{\text{comp}}$ is the computing signal for the pre-processed $l$-th local model parameter at the $k$-th sensor, which generates from AI model trained by its local dataset, and ${s}_{k,j}^{\text{comm}}$ is the communication signal for the $j$-th measured data at the $k$-th sensor, which needs to be decoded exactly at the BS.  For ease of analysis, it is assumed that the sensing signals, the computing signals, and the communication signals are mutually independent and all obey Gaussian distributed with unit norm, i.e., $\mathbb{E}\left\{ s_{k,i}^{\text{sens}}{{\left( s_{k,i}^{\text{sens}} \right)}^{H}} \right\}=\mathbb{E}\left\{ s_{k,l}^{\text{comp}}{{\left( s_{k,l}^{\text{comp}} \right)}^{H}} \right\}=\mathbb{E}\left\{ s_{k,j}^{\text{comm}}{{\left( s_{k,j}^{\text{comm}} \right)}^{H}} \right\}=1, \forall k\in\Omega_K,i\in\Omega_I,l\in\Omega_L,j\in\Omega_J$.  Then, the $k$-th sensor constructs a superposition coded transmit signal as
\begin{equation}\label{trans_sig}
  {{\mathbf{x}}_{k}}=\underbrace{\sum\limits_{i=1}^{I}{{{\mathbf{a}}_{k,i}}s_{k,i}^{\text{sens}}}}_{\text{Sensing}}+\underbrace{\sum\limits_{l=1}^{L}{{{\mathbf{b}}_{k,l}}s_{k,l}^{\text{comp}}}}_{\text{Computing}}+\underbrace{\sum\limits_{j=1}^{J}{{{\mathbf{c}}_{k,j}}s_{k,j}^{\text{comm}}}}_{\text{Communication}},
\end{equation}
where ${{\mathbf{a}}_{k,i}}\in {\mathbb{C}^{M\times \text{1}}},{{\mathbf{b}}_{k,l}}\in {\mathbb{C}^{M\times \text{1}}},{{\mathbf{c}}_{k,j}}\in {\mathbb{C}^{M\times \text{1}}}$ are the transmit beamforming vectors for corresponding sensing signal, computing signal, and communication signal, respectively. Consequently, the received signal at the BS is given by
\begin{eqnarray}\label{rece_sig}
  \mathbf{y}\!\!&=&\!\!\underbrace{\sum\limits_{k=1}^{K}{\sum\limits_{i=1}^{I}{{{r}_{i}}{{\mathbf{G}}_{k,i}}{{\mathbf{a}}_{k,i}}s_{k,i}^{\text{sens}}}}}_{\text{Sensing}}+\underbrace{\sum\limits_{k=1}^{K}{\sum\limits_{o=1}^{O}{{{r}_{o}}{{\mathbf{F}}_{k,o}}\sum\limits_{m=1}^{I}{{{\mathbf{a}}_{k,m}}s_{k,m}^{\text{sens}}}}}}_{\text{Interference from clutters}}\nonumber \\
  \!\!&+&\!\!\underbrace{\sum\limits_{k=1}^{K}{\sum\limits_{l=1}^{L}{{{\mathbf{H}}_{k}}{{\mathbf{b}}_{k,l}}s_{k,l}^{\text{comp}}}}}_{\text{Computing}}+
 \underbrace{\sum\limits_{k=1}^{K}{\sum\limits_{j=1}^{J}{{{\mathbf{H}}_{k}}{{\mathbf{c}}_{k,j}}s_{k,j}^{\text{comm}}}}}_{\text{Communication}}+\underbrace{\mathbf{n}}_{\text{Noise}},
\end{eqnarray}
where $\mathbf{n}$ is the additive white Gaussian noise (AWGN) vector, $r_{i} \in \mathbb{C}, \forall i\in\Omega_I,$ and $r_{o}\in \mathbb{C}, \forall o\in \Omega_O,$ are the reflection coefficients of the $i$-th target and the $o$-th clutter, respectively.
 ${\mathbf{G}_{k,i}}=\mathbf{g}_{k,i}^{'}\mathbf{g}_{k,i}^{H} \in \mathbb{C}^{N\times M}$ represents the cascaded channel in which $\mathbf{g}_{k,i}$ and $\mathbf{g}_{k,i}^{'}$ are the channel gains from the $k$-th sensor to the $i$-th target and from the $i$-th target to the BS, respectively. ${\mathbf{F}_{k,o}}=\mathbf{f}_{k,o}^{'}\mathbf{f}_{k,o}^{H}\in \mathbb{C}^{N\times M}$ denotes the cascaded channel, where $\mathbf{f}_{k,o}$ and $\mathbf{f}_{k,o}^{'}$ are the channel gains from the $k$-th sensor to the $o$-th clutter and from the $o$-th clutter to the BS, respectively. Furthermore, $\mathbf{H}_k\in \mathbb{C}^{N\times M}$ is the channel gain from the $k$-th sensor to the BS. Here, we assume that ${\mathbf{G}_{k,i}}$, ${\mathbf{F}_{k,o}}$, and ${{\mathbf{H}}_{k}}$ remain constant within a time slot, but fade over time slots independently. In the following, we will discuss the performance metrics of the three fundamental tasks one-by-one.

 %To effectively integrate sensing, computing and communication, it is desired to design the transceiver carefully according to their performance requirements. Hence, we select the appropriate performance metrics based on the functions of sensing, computing and communication. Specifically, the performance metric for sensing and computing is mean squared error (MSE) in order to characterize their accuracy, while the performance metric for communication is signal-to-interference-plus-noise ratio (SINR) in order to guarantee the quality of received signals. In the following, we will discuss the performance metrics of the three fundamental tasks one-by-one.

\subsection{Sensing for Target Observation}
\begin{figure}[!h]
 \centering
\includegraphics [width=0.45\textwidth] {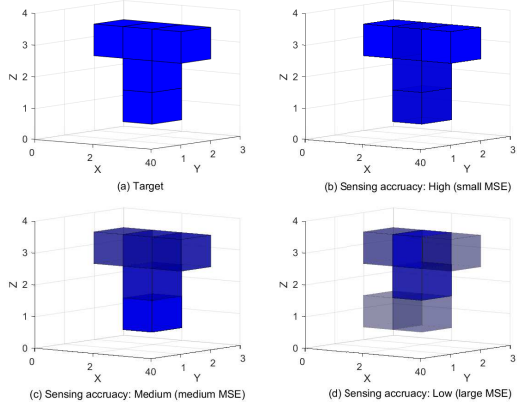}
\caption {Target Imaging results under different sensing accuracy.}
\label{Sensing_fig2}
\end{figure}

First, we deal with the sensing signal. To improve the accuracy of target sensing, all sensors collaborate to sense the targets. Thus, a linear unbiased estimator is deployed at the BS to estimate the reflection coefficients of the targets. To obtain an accurate reflection coefficient $r_{i}, i\in\Omega_I,$ for target sensing, the BS carries out receive beamforming to enhance the desired signal and suppress the co-channel interference. Hence, the estimated reflection coefficient ${{{\hat{r}}}_{i}}$ of the $i$-th target can be expressed as
\begin{eqnarray}\label{esti_ref}
\!\!{{{\hat{r}}}_{i}}\!\!\!\!&=&\!\!\!\!\mathbf{v}_{i}^{H}\mathbf{y}\nonumber\\
\!\!\!\!&=&\!\!\!\!\mathbf{v}_{i}^{H}{{r}_{i}}\sum\limits_{k=1}^{K}{{{\mathbf{G}}_{k,i}}{{\mathbf{a}}_{k,i}}s_{k,i}^{\text{sens}}}+\mathbf{v}_{i}^{H}\sum\limits_{k=1}^{K}{\sum\limits_{m=1,m\ne i}^{I}{{{r}_{m}}{{\mathbf{G}}_{k,m}}{{\mathbf{a}}_{k,m}}s_{k,m}^{\text{sens}}}}\nonumber\\
\!\!\!\!&+&\!\!\!\!\mathbf{v}_{i}^{H}\sum\limits_{k=1}^{K}{\sum\limits_{o=1}^{O}{{{r}_{o}}{{\mathbf{F}}_{k,o}}\sum\limits_{m=1}^{I}{{{\mathbf{a}}_{k,m}}s_{k,m}^{\text{sens}}}}}
\nonumber\\
\!\!\!\!&+&\!\!\!\!\mathbf{v}_{i}^{H}\sum\limits_{k=1}^{K}{{{\mathbf{H}}_{k}}\left( \sum\limits_{l=1}^{L}{{{\mathbf{b}}_{k,l}}s_{k,l}^{\text{comp}}}+\sum\limits_{j=1}^{J}{{{\mathbf{c}}_{k,j}}s_{k,j}^{\text{comm}}} \right)}\nonumber\\
\!\!\!\!&+&\!\!\!\!\mathbf{v}_{i}^{H}\mathbf{n}, \forall i,
\end{eqnarray}
where $\mathbf{v}_{i} \in \mathbb{C}^{N\times 1}$ is the sensing receive beamforming vector adopted at the BS for the $i$-th target. Following relevant parameter estimation works, we adopt the mean squared error (MSE) as the performance metric of target sensing.  It is seen from Eq. (\ref{esti_ref}) that the $\hat{r}_i$ is not only a function of $s_{k,i}^{\text{sens}}$, but also the other sensing signals from different sensors for different targets, the computing signals, and communication signals. In this context, we can minimize the MSE between the estimated reflection coefficient ${{{\hat{r}}}_{i}}$ and the actual reflection coefficient $r_{i}$ to suppress the interference from other signals and achieve a high-accuracy target sensing. Thus, the MSE is given by
\begin{eqnarray}
  \text{MSE}_{i}^{\text{sens}}&=&\mathbb{E}\left\{ \left( {{{\hat{r}}}_{i}}-{{r}_{i}} \right){{\left( {{{\hat{r}}}_{i}}-{{r}_{i}} \right)}^{H}} \right\} \nonumber\\
  &=&{{\left| \sum\limits_{k=1}^{K}{\mathbf{v}_{i}^{H}{{\mathbf{G}}_{k,i}}{{\mathbf{a}}_{k,i}}}-1 \right|}^{2}}R_{\text{i}}^{2} \nonumber\\
  &+&\sum\limits_{k=1}^{K}{\sum\limits_{m=1,m\ne i}^{I}{R_{\text{m}}^{2}{{\left| \mathbf{v}_{i}^{H}{{\mathbf{G}}_{k,m}}{{\mathbf{a}}_{k,m}} \right|}^{2}}}}\nonumber\\
  &+&\sum\limits_{k=1}^{K}{\sum\limits_{o=1}^{O}{\sum\limits_{m=1}^{I}{R_{o}^{2}{{\left| \mathbf{v}_{i}^{H}{{\mathbf{F}}_{k,o}}{{\mathbf{a}}_{k,m}} \right|}^{2}}}}} \nonumber \\
  &+&\sum\limits_{k=1}^{K}{\left( \sum\limits_{l=1}^{L}{{{\left| \mathbf{v}_{i}^{H}{{\mathbf{H}}_{k}}{{\mathbf{b}}_{k,l}} \right|}^{2}}}+ \sum\limits_{j=1}^{J}{{{\left| \mathbf{v}_{i}^{H}{{\mathbf{H}}_{k}}{{\mathbf{c}}_{k,j}} \right|}^{2}}} \right)} \nonumber \\
  &+&\sigma _{n}^{2}{{\left\| \mathbf{v}_{i}^{H} \right\|}^{2}}, \forall i,\label{MSE_sens}
\end{eqnarray}
where $\sigma_n^2$ is the noise power, and ${{R}_{m}}$ is the root mean squared (RMS) value of the reflection coefficient ${{r}_{m}}, \forall m\in\{\Omega_I\cup\Omega_O\}$ according to the a-priori probability of occurrence for each type \cite{target}.  To present more intuitively the impact of reflection coefficient estimation, we provide an application result on target imaging \cite{imaging}. As is shown in Fig. \ref{Sensing_fig2}, the higher the sensing accuracy (the smaller MSE between the estimated reflection coefficient and the actual reflection coefficient), the more accurate the target imaging.

\subsection{Computing for Model Aggregation}
Secondly, we handle the computing signal.  In a time slot, a shared learning process for AirFL with a global model $\mathbf{s}^{\text{comp}}\in \mathbb{R}^{L\times1}$ is trained cooperatively by multiple intelligent sensors \cite{ICC7,ICC8}. Each intelligent sensor has its own local dataset $\mathcal{D}_{k},\forall k\in\Omega_K$, and the sum of local datasets is $\mathcal{D}$, where $\left| \mathcal{D} \right|=\sum\limits_{k=1}^{K}{\left| {{\mathcal{D}}_{k}} \right|}$. Thus, the desired global model at the BS by computing local models is given by
\begin{equation}\label{aggre_model}
s_{l}^{\text{comp}}=\frac{1}{K}\sum\limits_{k=1}^{K}{s_{k,l}^{\text{comp}}}, \forall l \in \Omega_L
\end{equation}
where $s_{k,l}^{\text{comp}}$ is the $l$-th computing signal contributed by the pre-processed local model parameter at the $k$-th sensor, and each local model parameter is multiplied by a weighting factor ${{\xi }_{k}}=\frac{\left| {{\mathcal{D}}_{k}} \right|}{\left| \mathcal{D} \right|}$ before transmission. To reduce the computation error in the presence of interference and channel fading, we need to implement receive beamforming at the BS. As a result, the received signal for AirFL at the BS can be expressed as
\begin{eqnarray}
\hat{s}_{l}^{\text{comp}}\!\!\!\!&=&\!\!\!\!\frac{\mathbf{z}_{l}^{H}}{K}\mathbf{y}\nonumber\\
\!\!\!\!&=&\!\!\!\!\frac{\mathbf{z}_{l}^{H}}{K}\sum\limits_{k=1}^{K}{{{\mathbf{H}}_{k}}{{\mathbf{b}}_{k,l}}s_{k,l}^{\text{comp}}+\frac{\mathbf{z}_{l}^{H}}{K}\sum\limits_{k=1}^{K}{\sum\limits_{i=1,i\ne l}^{L}{{{\mathbf{H}}_{k}}{{\mathbf{b}}_{k,i}}s_{k,i}^{\text{comp}}}}}\nonumber\\
\!\!\!\!&+&\!\!\!\!\frac{\mathbf{z}_{l}^{H}}{K}\sum\limits_{k=1}^{K}{\sum\limits_{i=1}^{I}{{{r}_{i}}{{\mathbf{G}}_{k,i}}{{\mathbf{a}}_{k,i}}s_{k,i}^{\text{sens}}}} \nonumber \\
\!\!\!\!&+&\!\!\!\!\frac{\mathbf{z}_{l}^{H}}{K}\sum\limits_{k=1}^{K}{\sum\limits_{o=1}^{O}{{{r}_{o}}{{\mathbf{F}}_{k,o}}\sum\limits_{m=1}^{I}{{{\mathbf{a}}_{k,m}}s_{k,m}^{\text{sens}}}}}\nonumber\\
\!\!\!\!&+&\!\!\!\!\frac{\mathbf{z}_{l}^{H}}{K}\sum\limits_{k=1}^{K}{\sum\limits_{j=1}^{J}{{{\mathbf{H}}_{k}}{{\mathbf{c}}_{k,j}}s_{k,j}^{\text{comm}}}}+\frac{\mathbf{z}_{l}^{H}}{K}\mathbf{n},\forall l,
\end{eqnarray}
where $\mathbf{z}_{l} \in \mathbb{C}^{N\times 1}$ is the computing receive beamforming vector at the BS for the $l$-th model parameter.  In general, the distortion of the model aggregation is measured by the MSE between ${s}_l^{\text{comp}}$ and ${\hat{s}_l^{\text{comp}}}$, which can be expressed as
\begin{eqnarray}
\!\!\!\text{MSE}_{l}^{\text{comp}}\!\!\!\!&=&\!\!\!\!\mathbb{E}\left\{\left( \hat{s}_{l}^{\text{comp}}-s_{l}^{\text{comp}} \right){{\left( \hat{s}_{l}^{\text{comp}}-s_{l}^{\text{comp}} \right)}^{H}} \right\}\nonumber\\
    \!\!\!\!&=&\!\!\!\!\frac{1}{{{K}^{2}}}\sum\limits_{k=1}^{K}{{{\left| \mathbf{z}_{l}^{H}{{\mathbf{H}}_{k}}{{\mathbf{b}}_{k,l}}-1 \right|}^{2}}} \nonumber\\
    \!\!\!\!&+&\!\!\!\!\frac{1}{{{K}^{2}}}\sum\limits_{m=1,m\ne l}^{L}{\sum\limits_{k=1}^{K}{{{\left| \mathbf{z}_{l}^{H}{{\mathbf{H}}_{k}}{{\mathbf{b}}_{k,m}} \right|}^{2}}}}\nonumber\\
    \!\!\!\!&+&\!\!\!\!\frac{1}{{{K}^{2}}}\sum\limits_{k=1}^{K}{\sum\limits_{i=1}^{I}{R_{i}^{2}{{\left| \mathbf{z}_{l}^{H}{{\mathbf{G}}_{k,i}}{{\mathbf{a}}_{k,i}} \right|}^{2}}}}\nonumber\\
    \!\!\!\!&+&\!\!\!\!\frac{1}{{{K}^{2}}}\sum\limits_{k=1}^{K}{\sum\limits_{o=1}^{O}{\sum\limits_{m=1}^{I}{R_{o}^{2}{{\left| \mathbf{z}_{l}^{H}{{\mathbf{F}}_{k,o}}{{\mathbf{a}}_{k,m}} \right|}^{2}}}}}\nonumber\\
    \!\!\!\!&+&\!\!\!\!\frac{1}{{{K}^{2}}}\sum\limits_{k=1}^{K}{\sum\limits_{j=1}^{J}{{{\left| \mathbf{z}_{l}^{H}{{\mathbf{H}}_{k}}{{\mathbf{c}}_{k,j}} \right|}^{2}}}}+\frac{\sigma _{n}^{2}}{{{K}^{2}}}{{\left\| {{\mathbf{z}}_{l}} \right\|}^{2}}, \forall l.\label{MSE_comp}
\end{eqnarray}
 Herein, we also provide a classical application case to vividly present the function of computing, i.e. image recognition based on AirFL\cite{digit}. It is seen from Table \ref{digit2} that the higher computing accuracy (the smaller computing MSE between the aggregated computing signal and the desired global model), the more accurate the image recognition.

\begin{table*}
  \caption{Handwritten-digit identification results under different computing accuracy on IID MNIST dataset trained by CNN network}
  \label{digit2}
  \centering
  \tiny
    \begin{tabular}{ | c | c | c | c | c| c |c |c | }
    \hline
    Data image
    &\begin{minipage}[b]{0.07\columnwidth}
		\centering
		\raisebox{-.2\height}{\includegraphics[width=\linewidth]{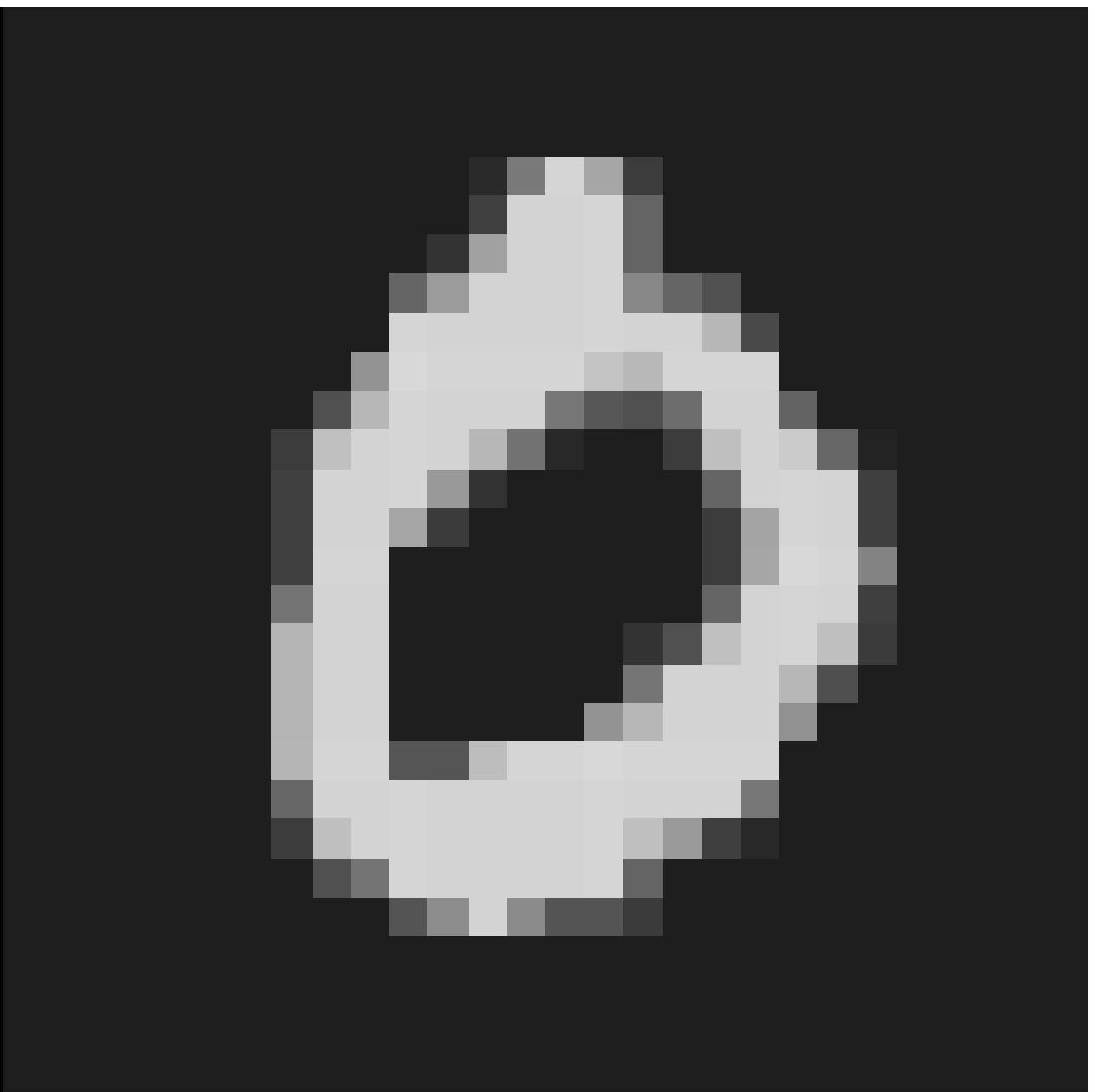}}
	\end{minipage}
    &     \begin{minipage}[b]{0.07\columnwidth}
		\centering
		\raisebox{-.2\height}{\includegraphics[width=\linewidth]{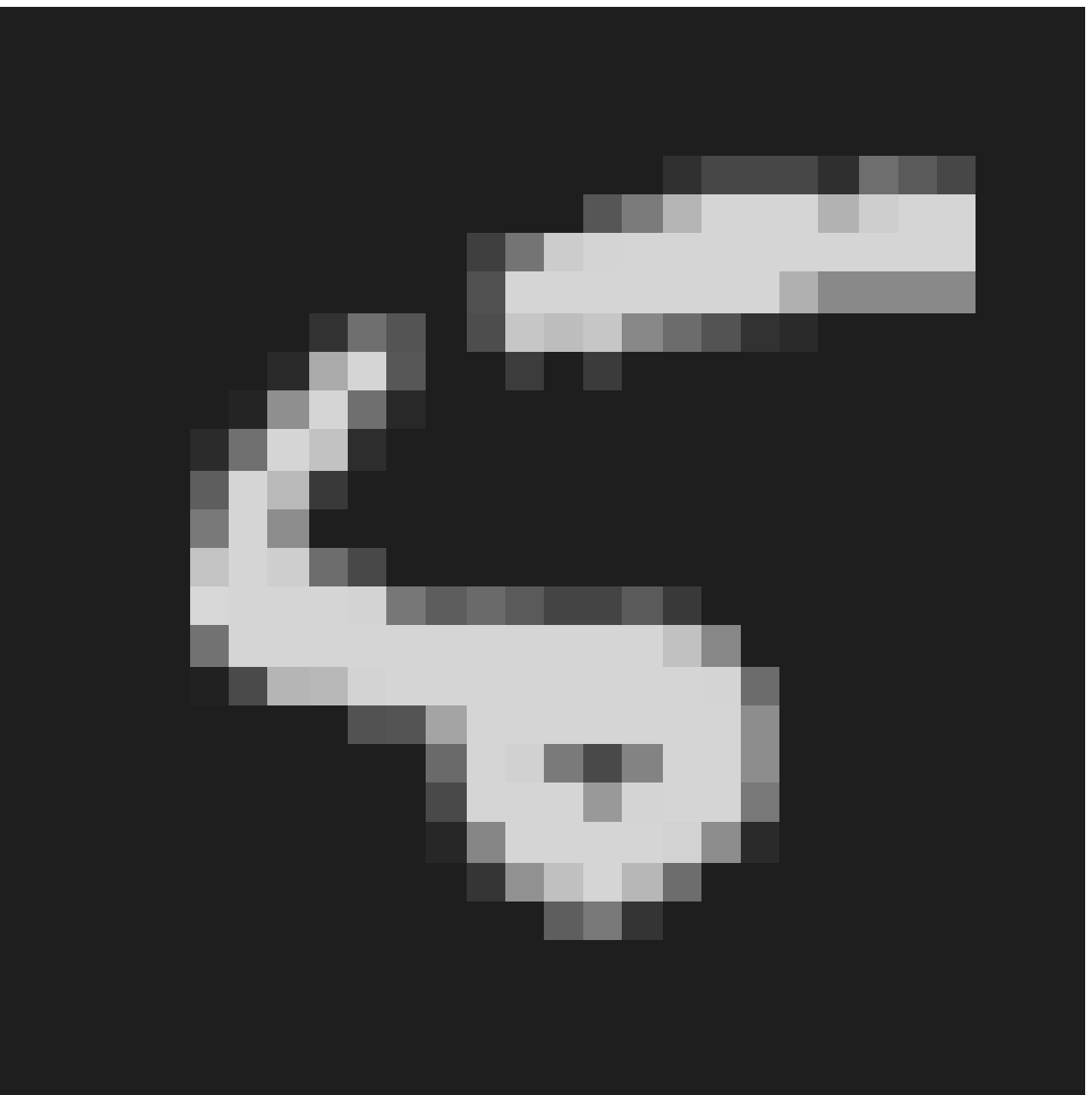}}
	\end{minipage}
    & \begin{minipage}[b]{0.07\columnwidth}
		\centering
		\raisebox{-.2\height}{\includegraphics[width=\linewidth]{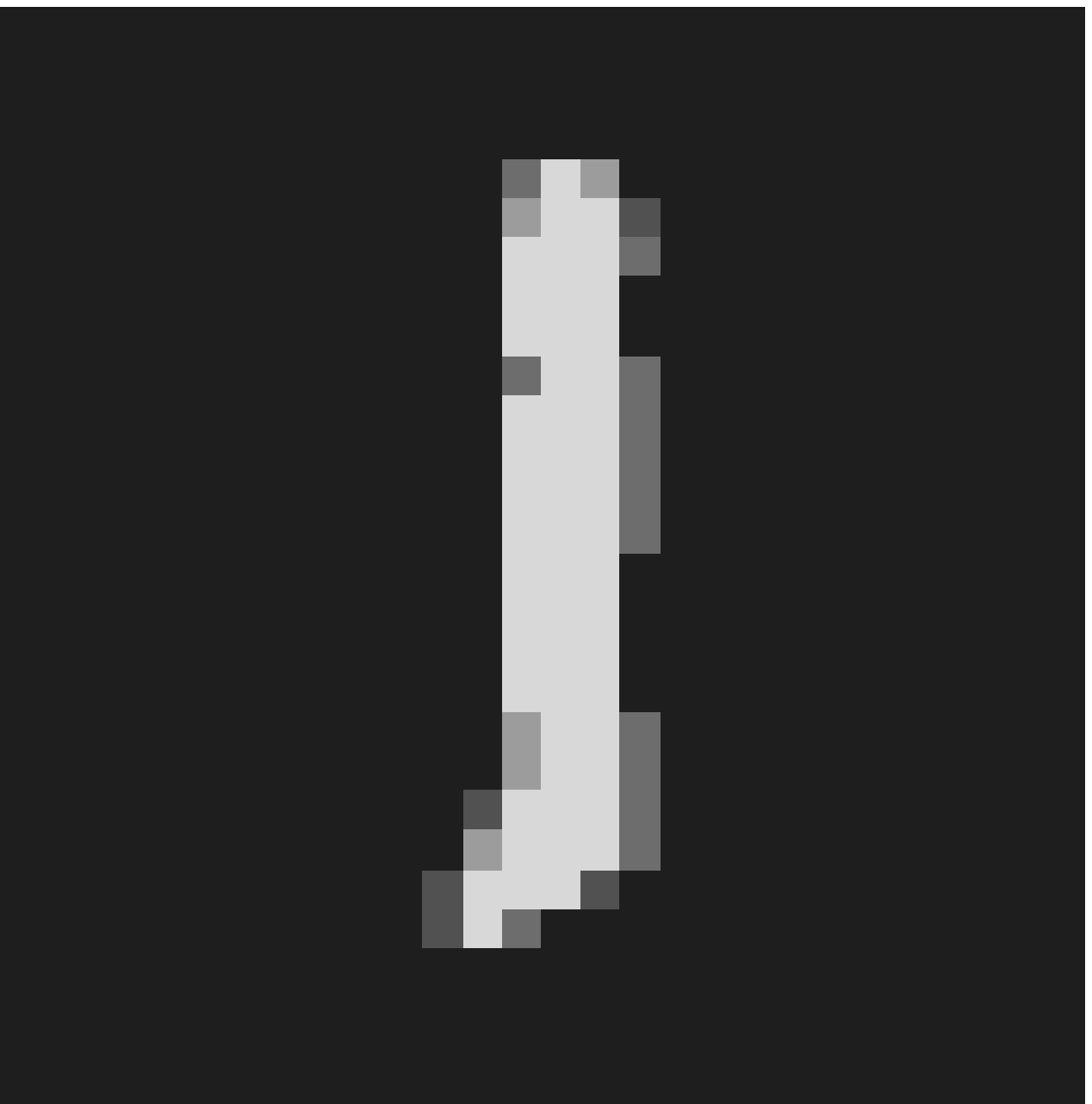}}
	\end{minipage}
& \begin{minipage}[b]{0.07\columnwidth}
		\centering
		\raisebox{-.2\height}{\includegraphics[width=\linewidth]{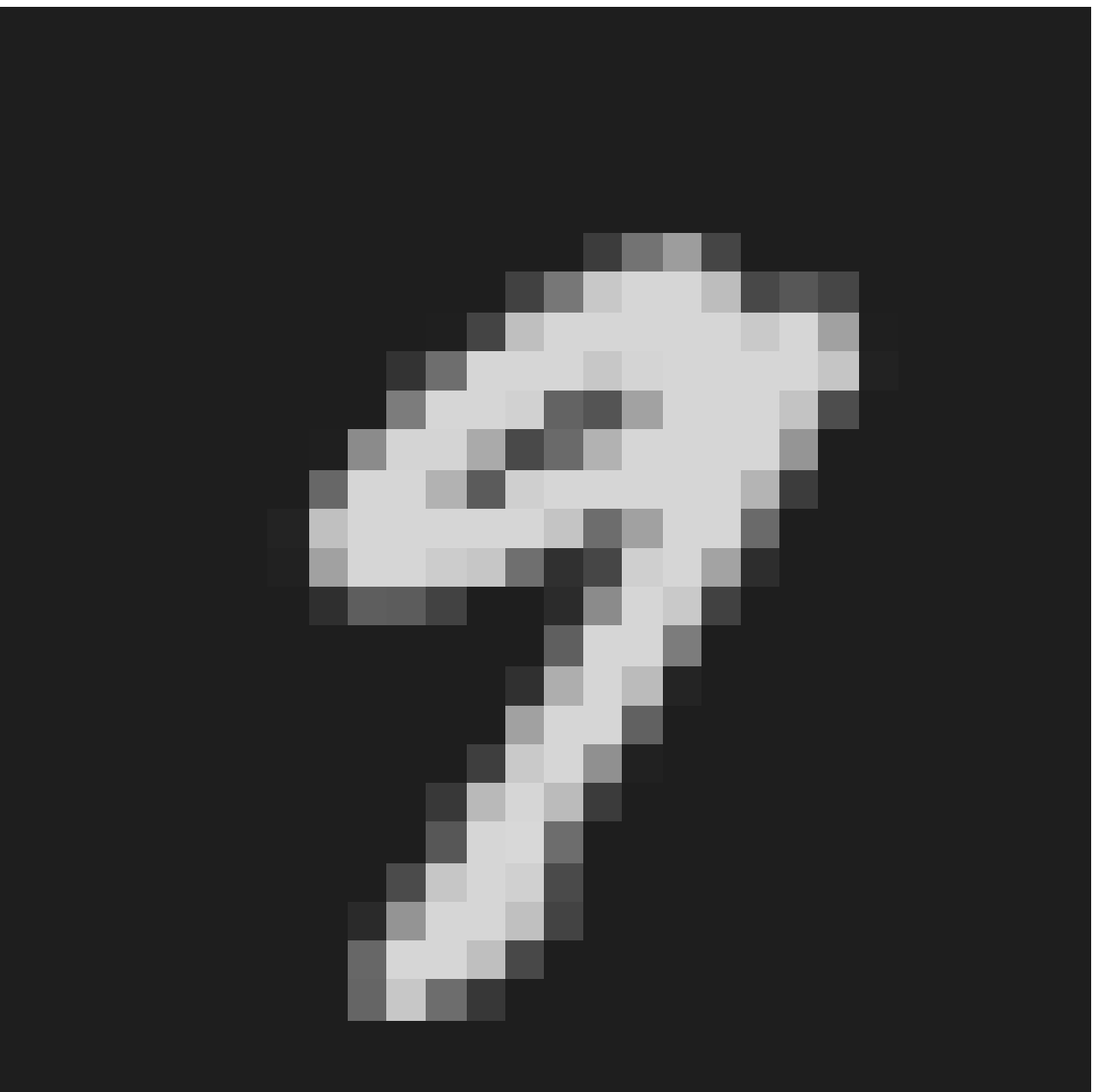}}
	\end{minipage}
& \begin{minipage}[b]{0.07\columnwidth}
		\centering
		\raisebox{-.2\height}{\includegraphics[width=\linewidth]{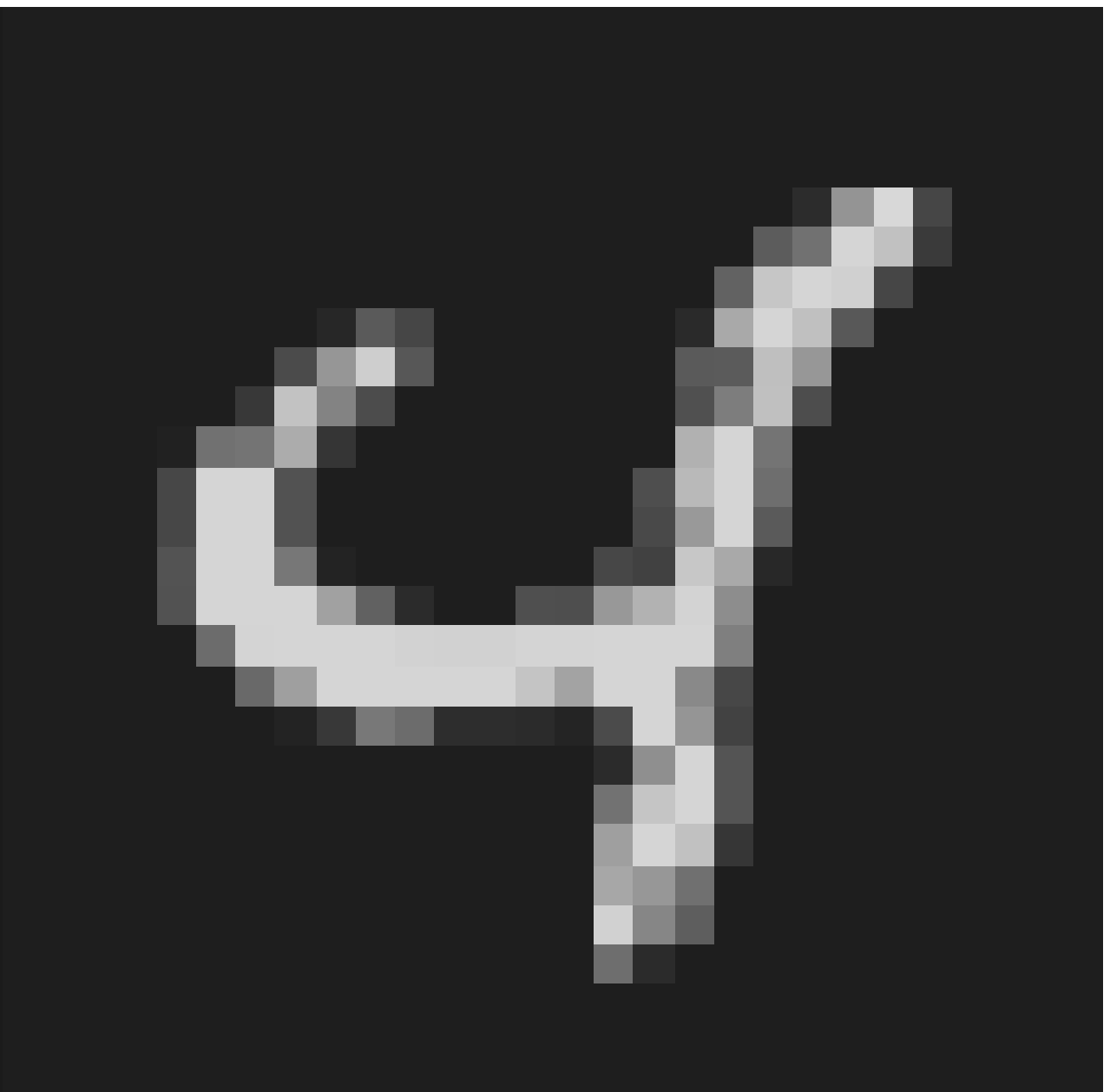}}
	\end{minipage}
& \begin{minipage}[b]{0.07\columnwidth}
		\centering
		\raisebox{-.2\height}{\includegraphics[width=\linewidth]{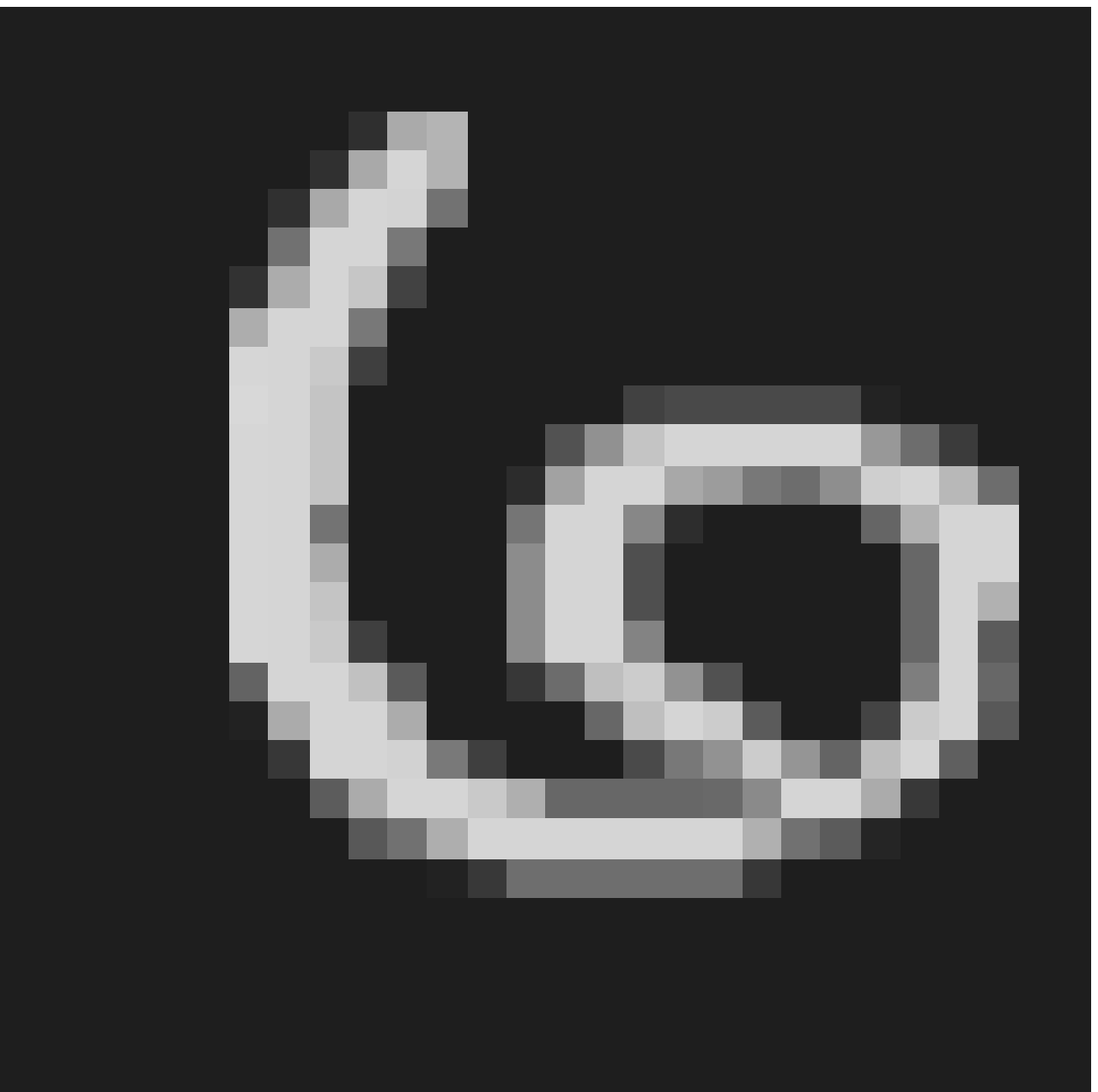}}
	\end{minipage}
& \begin{minipage}[b]{0.07\columnwidth}
		\centering
		\raisebox{-.2\height}{\includegraphics[width=\linewidth]{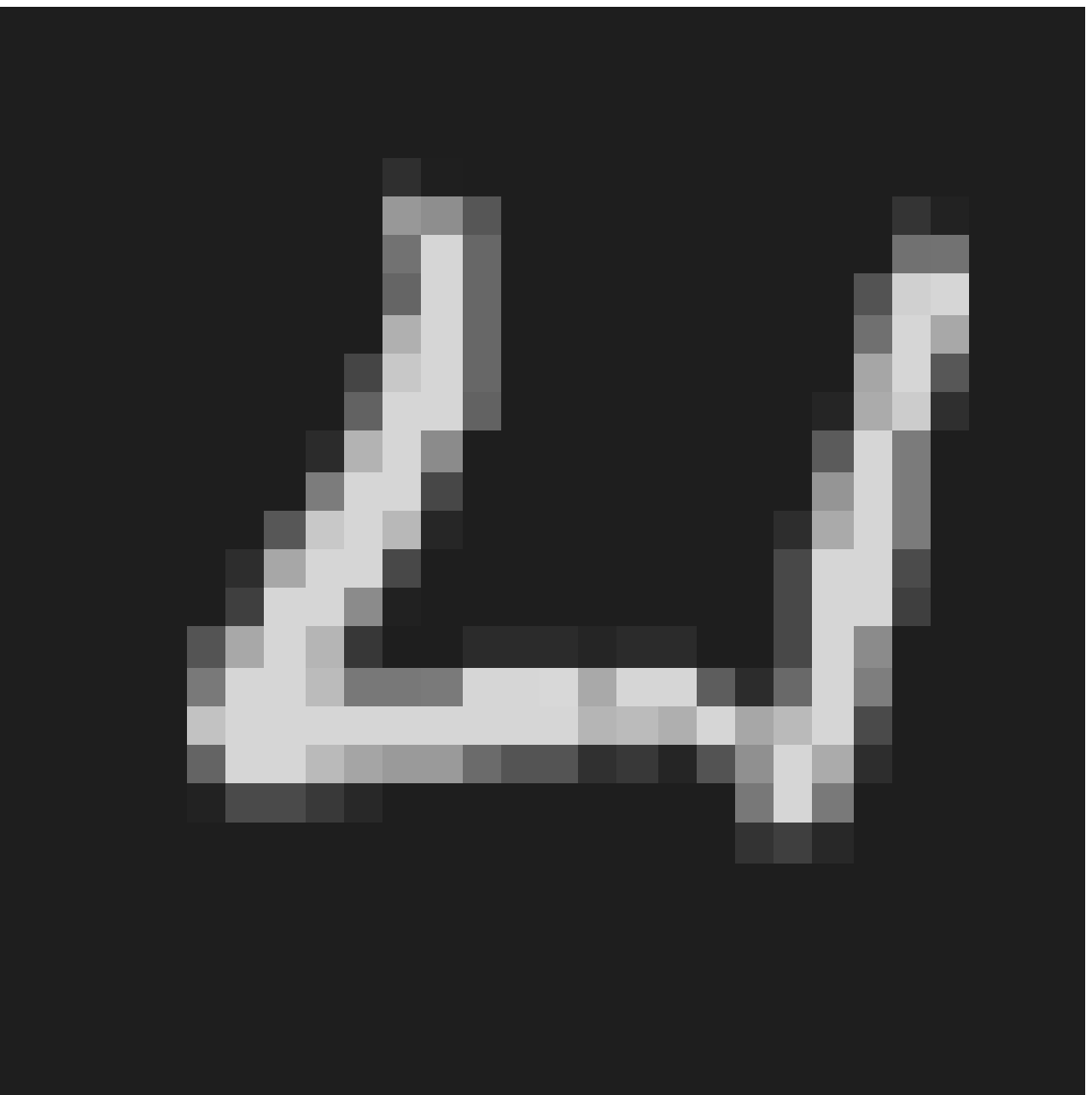}}
	\end{minipage}
\\ \hline
Computing accuracy: High (small MSE)  &0 &5 & 1 &9 &4 & 6 &4   \\ \hline
Computing accuracy: Medium (medium MSE)&6($\times$) &5& 1& 9& 4& 6& 4 \\ \hline
Computing accuracy: Low (large MSE) &0 &8($\times$) &1& 4($\times$)& 8($\times$)& 4($\times$) &4 \\ \hline\hline
Data image
& \begin{minipage}[b]{0.07\columnwidth}
		\centering
		\raisebox{-.2\height}{\includegraphics[width=\linewidth]{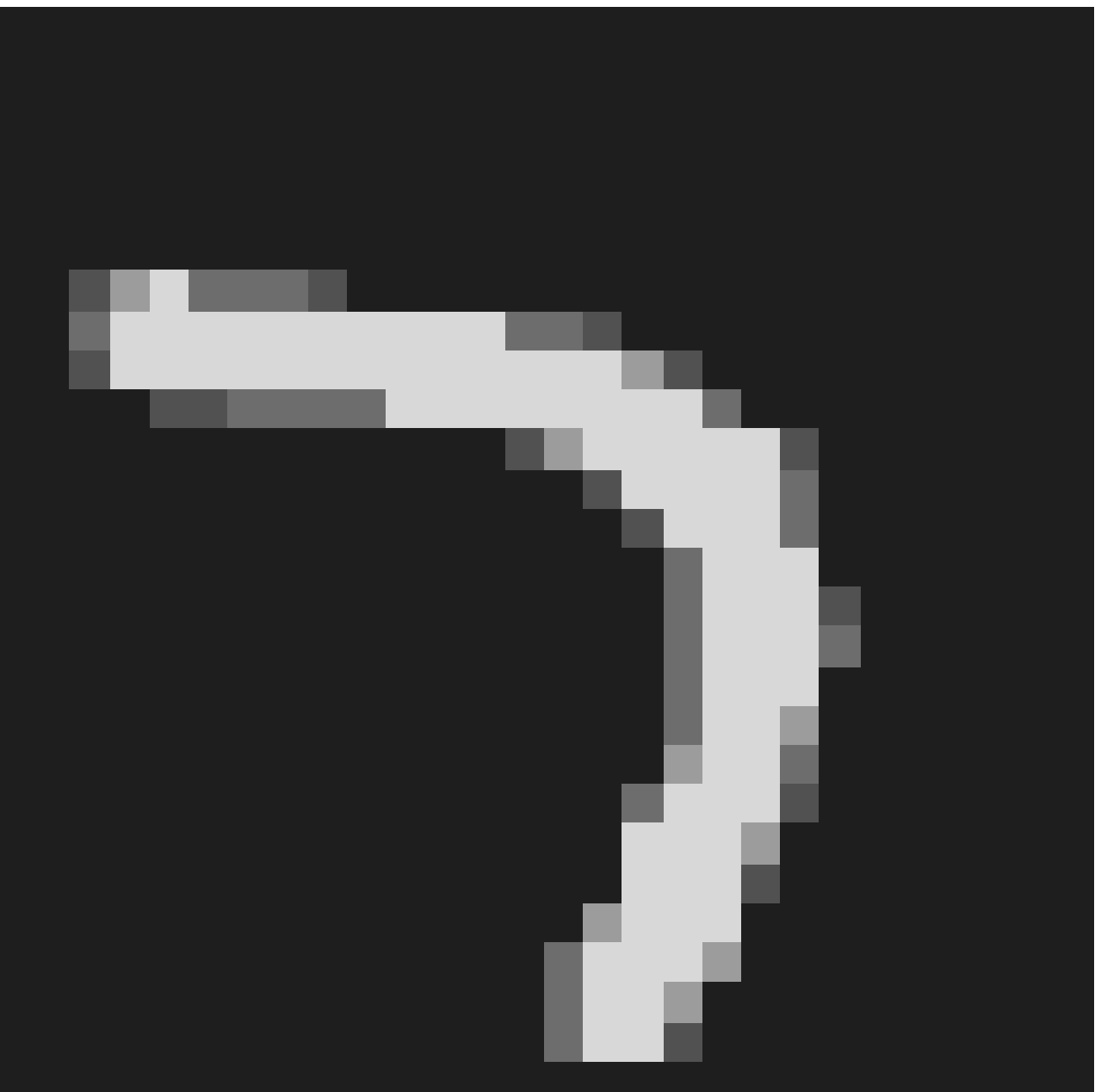}}
	\end{minipage}
& \begin{minipage}[b]{0.07\columnwidth}
		\centering
		\raisebox{-.2\height}{\includegraphics[width=\linewidth]{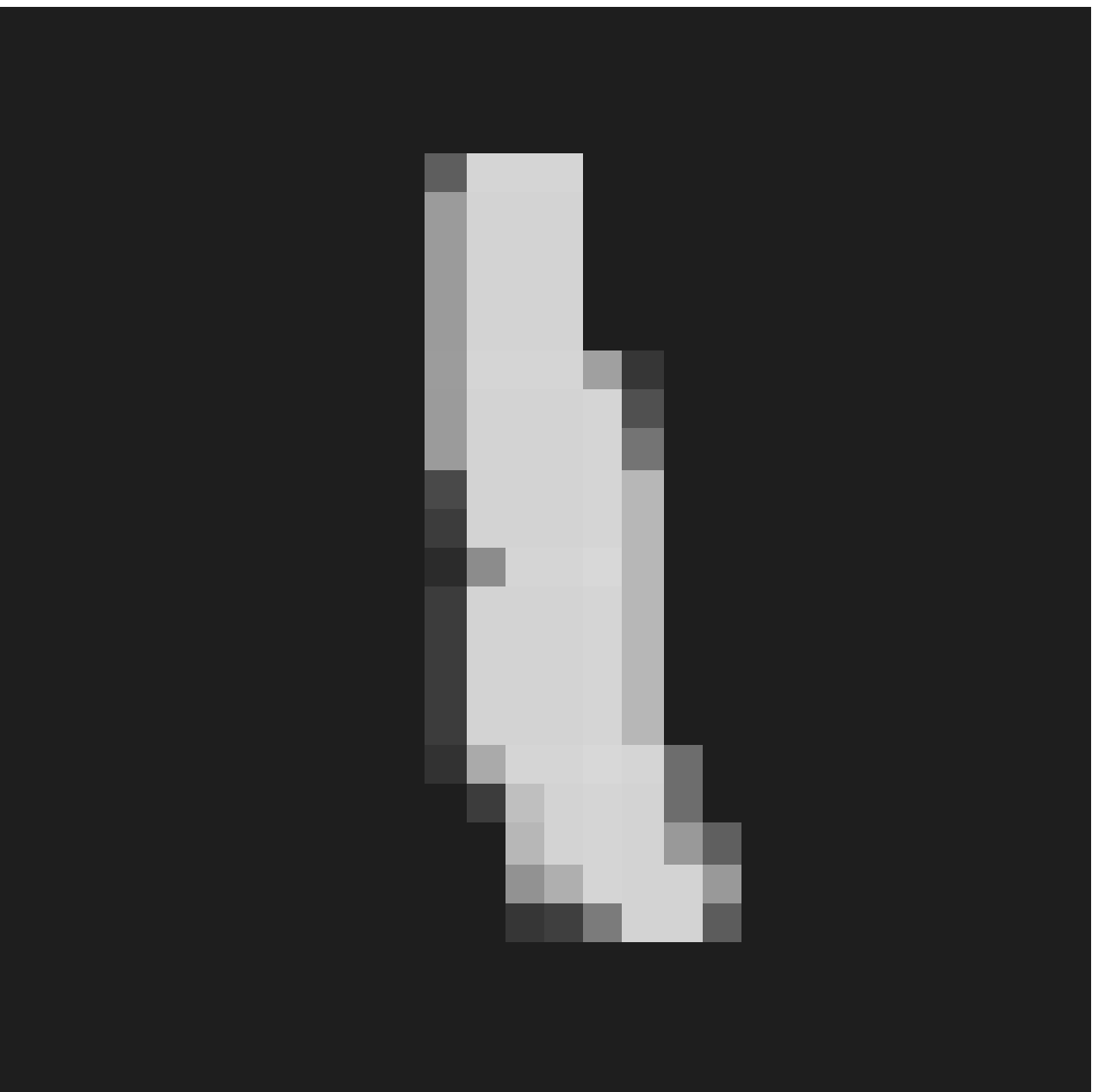}}
	\end{minipage}
& \begin{minipage}[b]{0.07\columnwidth}
		\centering
		\raisebox{-.2\height}{\includegraphics[width=\linewidth]{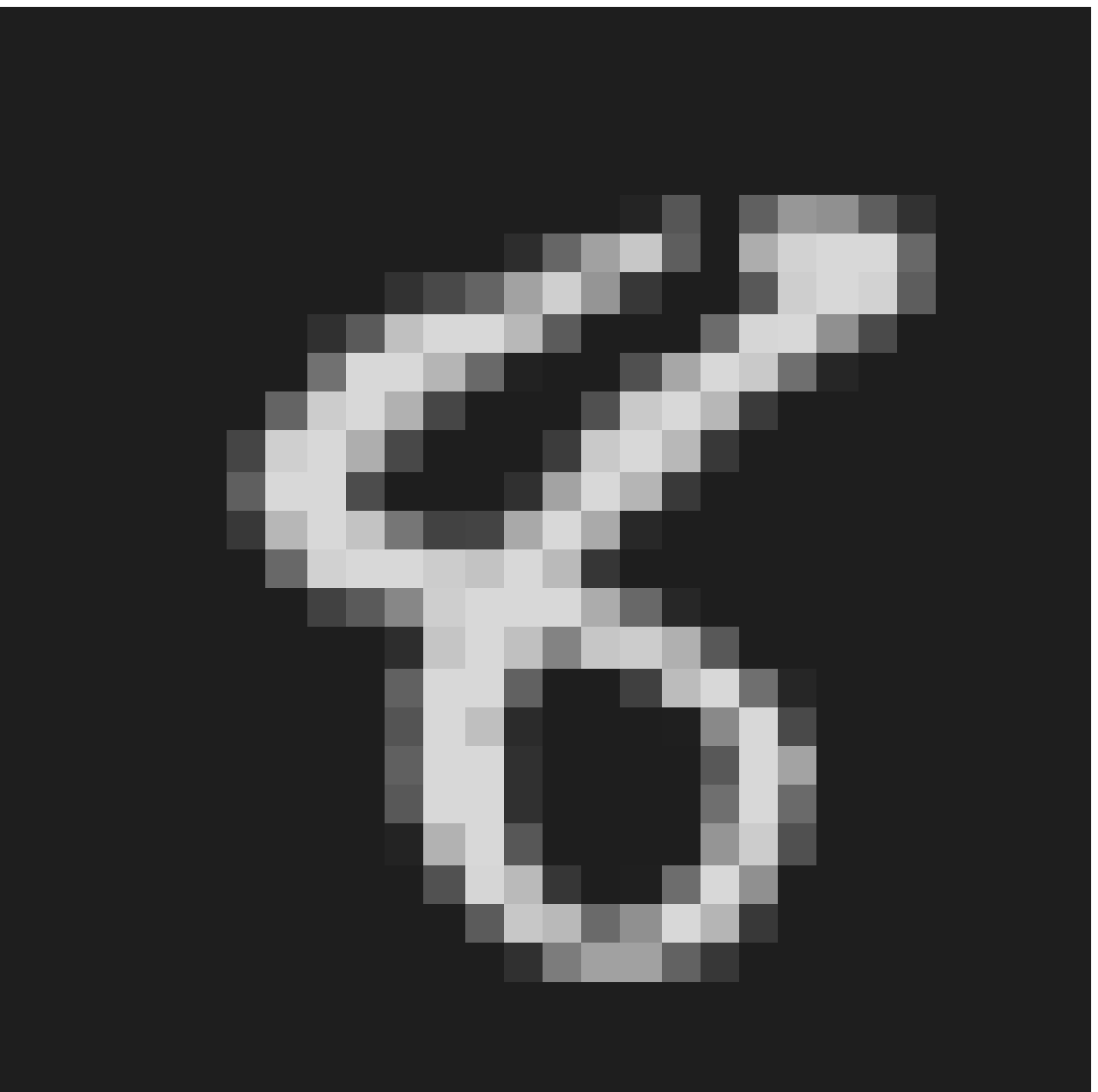}}
	\end{minipage}
& \begin{minipage}[b]{0.07\columnwidth}
		\centering
		\raisebox{-.2\height}{\includegraphics[width=\linewidth]{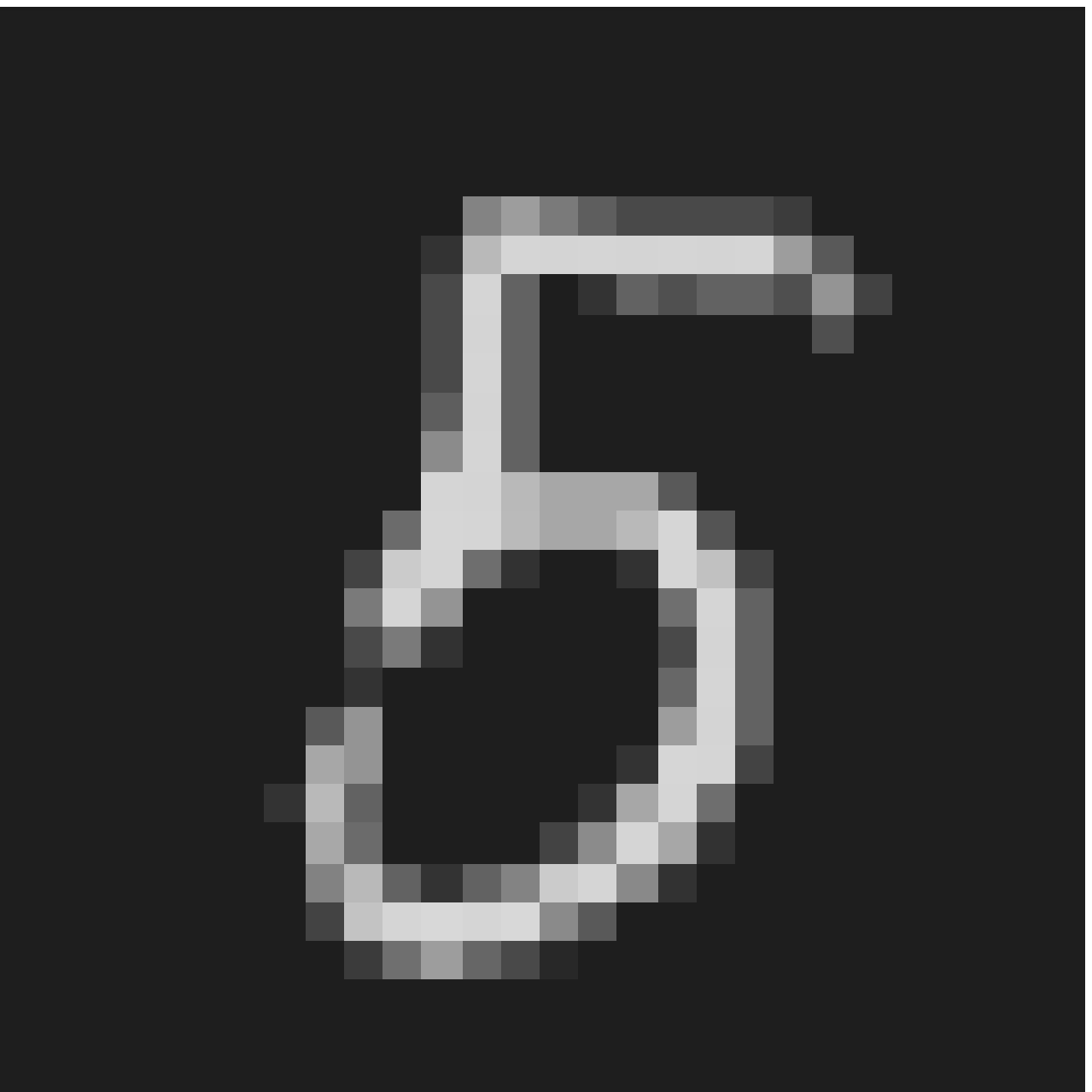}}
	\end{minipage}
& \begin{minipage}[b]{0.07\columnwidth}
		\centering
		\raisebox{-.2\height}{\includegraphics[width=\linewidth]{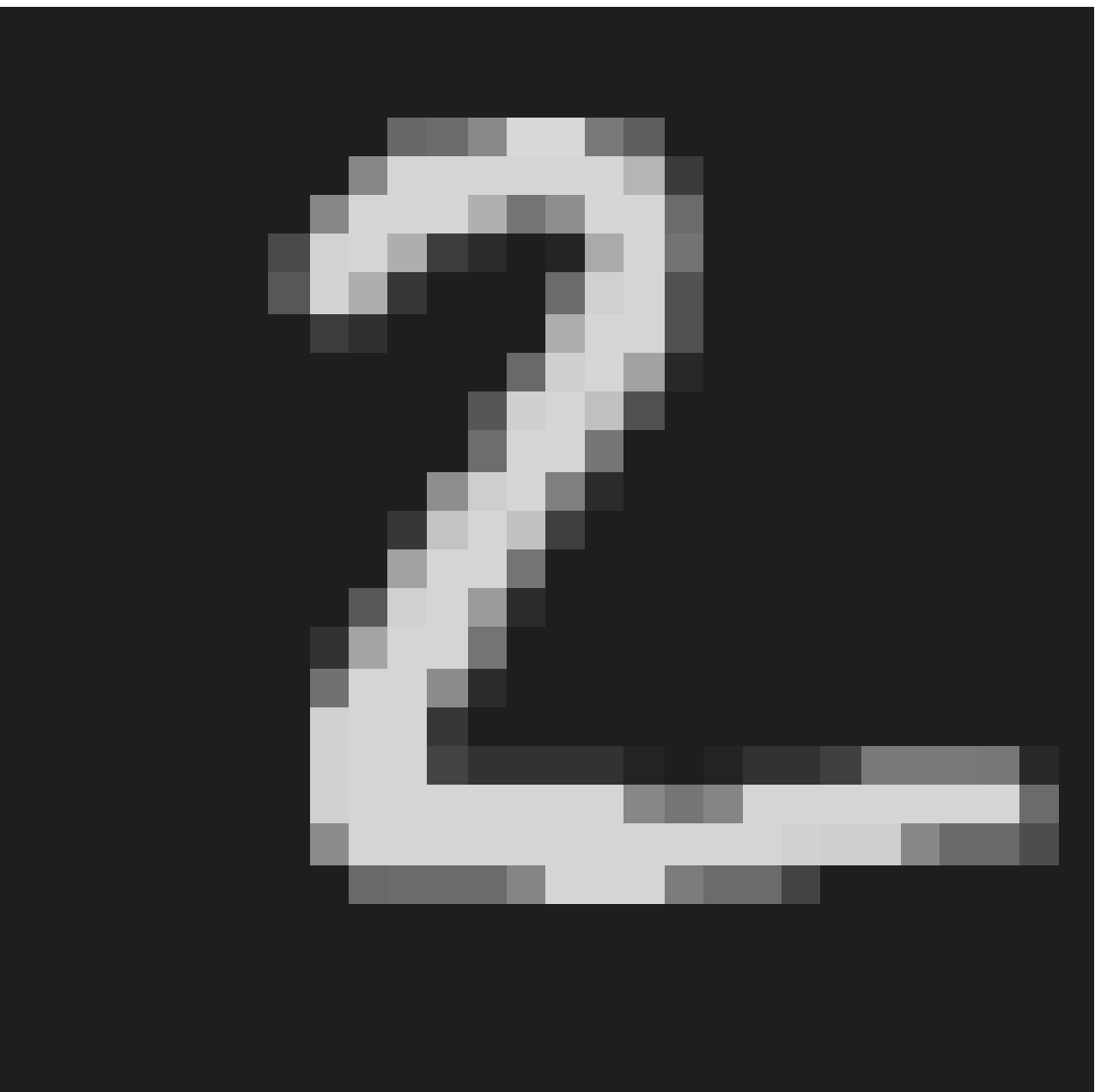}}
	\end{minipage}
& \begin{minipage}[b]{0.07\columnwidth}
		\centering
		\raisebox{-.2\height}{\includegraphics[width=\linewidth]{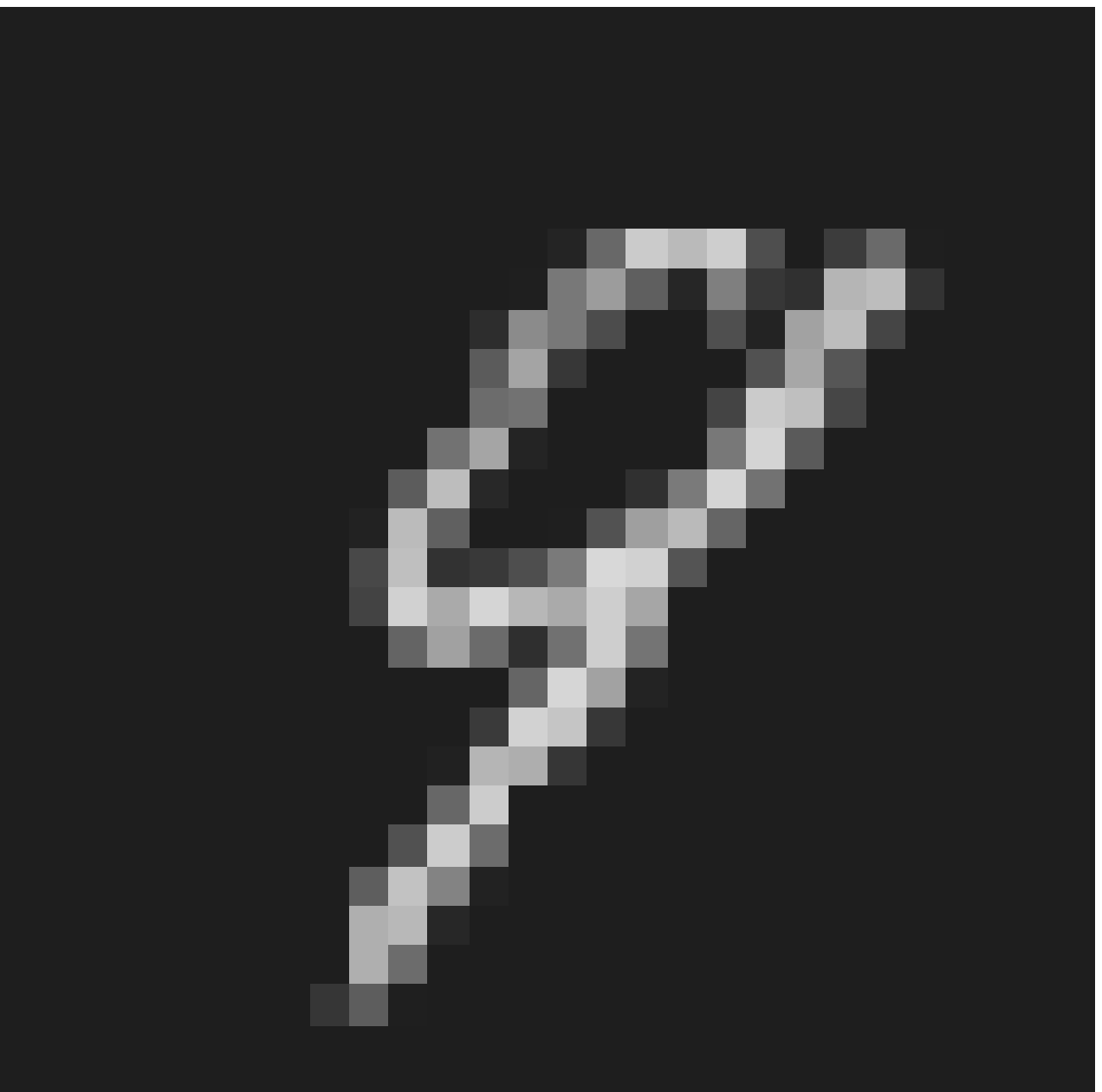}}
	\end{minipage}
& \begin{minipage}[b]{0.07\columnwidth}
		\centering
		\raisebox{-.2\height}{\includegraphics[width=\linewidth]{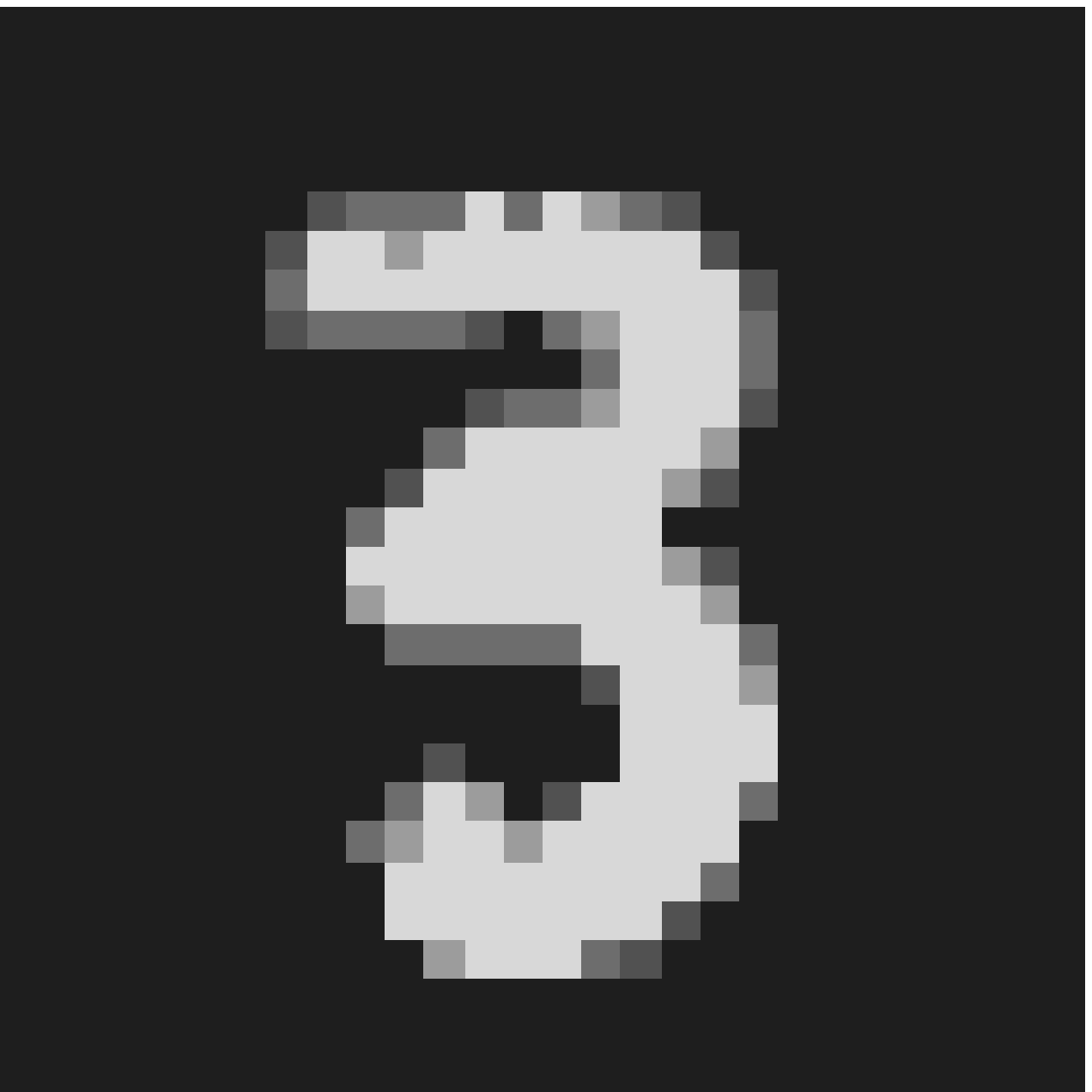}}
	\end{minipage}
    \\ \hline
Computing accuracy: High (small MSE)&7 &1&  8&  5& 2& 9& 3    \\ \hline
Computing accuracy: Medium (medium MSE)&7 &1&  8&  5& 2& 4($\times$)& 3 \\ \hline
Computing accuracy: Low (large MSE)& 4($\times$)& 1&8& 8($\times$)& 2 &4($\times$)& 8($\times$)  \\ \hline
  \end{tabular}
\end{table*}

\subsection{Communication for Information Transmission}
Finally, we discuss the communication signal. To promote the quality of the communication signal, the BS performs corresponding receive beamforming to mitigate the interference. Hence, the received signal at the BS for the communication signal $s_{k,j}^{\text{comm}}, \forall k\in\Omega_K,j\in\Omega_J,$ is given by
\begin{eqnarray}\label{rec_comm}
 y_{k,j}^{\text{comm}}&=&\mathbf{u}_{k,j}^{H}\mathbf{y} \nonumber\\
 &=&\mathbf{u}_{k,j}^{H}{{\mathbf{H}}_{k}}{{\mathbf{c}}_{k,j}}s_{k,j}^{\text{comm}} \nonumber\\
 &+&\sum\limits_{i=1,i\ne k}^{K}{\sum\limits_{n=1,n\ne j}^{J}{\mathbf{u}_{k,j}^{H}{{\mathbf{c}}_{i,n}}s_{i,n}^{\text{comm}}}} \nonumber\\
 &+&\sum\limits_{i=1}^{K}{\sum\limits_{l=1}^{L}{\mathbf{u}_{k,j}^{H}{{\mathbf{H}}_{i}}{{\mathbf{b}}_{i,l}}s_{i,l}^{\text{comp}}}}\nonumber\\
&+&\mathbf{u}_{k,j}^{H}\sum\limits_{k=1}^{K}{\sum\limits_{i=1}^{I}{{{r}_{i}}{{\mathbf{G}}_{k,i}}{{\mathbf{a}}_{k,i}}s_{k,i}^{\text{sens}}}}\nonumber\\
&+&\mathbf{u}_{k,j}^{H}\sum\limits_{k=1}^{K}{\sum\limits_{o=1}^{O}{{{r}_{o}}{{\mathbf{F}}_{k,o}}\sum\limits_{m=1}^{I}{{{\mathbf{a}}_{k,m}}s_{k,m}^{\text{sens}}}}}+\mathbf{u}_{k,j}^{H}\mathbf{n},
\end{eqnarray}
where $\mathbf{u}_{k,j} \in \mathbb{C}^{N\times 1}$ is the communication receive beamforming vector of the $j$-th communication signal from the $k$-th sensor. Since the received SINR determines the quality of the communication signal, it is usually regarded as the performance metric for communication. The signal-to-interference-plus-noise ratio (SINR) related to $s_{k,j}^{\text{comm}}$ can be expressed as
\begin{equation}\label{SINR}
{{\Gamma }_{k,j}}=\frac{{{\left| \mathbf{u}_{k,j}^{H}{{\mathbf{H}}_{k}}{{\mathbf{c}}_{k,j}} \right|}^{2}}}{\sum\limits_{i=1,i\ne k}^{K}{\sum\limits_{n=1,n\ne j}^{J}{{{\left| \mathbf{u}_{k,j}^{H}{{\mathbf{H}}_{i}}{{\mathbf{c}}_{i,n}} \right|}^{2}}}}+{{X}_{k,j}}+\sigma _{n}^{2}{{\left\| \mathbf{u}_{k,j}^{{}} \right\|}^{2}}},
\end{equation}
where
 \begin{eqnarray}
 {{X}_{k,j}}\!\!\!&=&\!\!\!\sum\limits_{i=1}^{K}{\sum\limits_{l=1}^{L}{{{\left| \mathbf{u}_{k,j}^{H}{{\mathbf{H}}_{i}}{{\mathbf{b}}_{i,l}} \right|}^{2}}}}+\sum\limits_{i=1}^{K}{\sum\limits_{m=1}^{I}{R_{m}^{2}{{\left| \mathbf{u}_{k,j}^{H}{{\mathbf{G}}_{i,m}}{{\mathbf{a}}_{i,m}} \right|}^{2}}}}\nonumber\\
 \!\!\!&+&\!\!\!\sum\limits_{i=1}^{K}{\sum\limits_{o=1}^{O}{\sum\limits_{m=1}^{I}{R_{o}^{2}{{\left| \mathbf{u}_{k,j}^{H}{{\mathbf{F}}_{i,o}}{{\mathbf{a}}_{i,m}} \right|}^{2}}}}}.
 \end{eqnarray}

It is observed from (\ref{MSE_sens}), (\ref{MSE_comp}), and (\ref{SINR}) that the performance of sensing, computing, and communication are jointly determined by the transmit beamforming vectors $\mathbf{a}_{k,i}$, $\mathbf{b}_{k,l}$, and $\mathbf{c}_{k,j}$ at the sensors, and receive beamforming vectors $\mathbf{v}_{i}$, $\mathbf{z}_{l}$, and $\mathbf{u}_{k,j}$ at the BS. Although the three performance metrics are all desirable to the system, they are competitive for the system resources. Hence, it is desired to establish a joint beamforming design framework to enhance the overall performance of ISCC over the limited resources in 6G wireless networks.

\emph{Remark:}  It is worthy pointing out that our proposed integrated system is inherently different from multi-user communication systems in terms of signal waveform, transceiver design and function realization. Firstly, multi-user communication systems only consider one kind of signal for communication purpose, while in our proposed integrated system, sensing, computing and communication signals have different types and serve different purposes. Secondly, signals of the same type from different users interfere with each other at the BS in multi-user communication systems, and thus transceiver is designed to reduce the inter-user interference.  In contrary,  for the proposed integrated system, signals of different types interfere with each other, but signals of the same type may help each other. In this case, the designed transceiver not only needs to decrease the interference among signals of different types, but also needs to increase the cooperative gain of signals of the same type in terms of sensing and computing. Finally, although different performance metrics such as SINR, data rate and MSE, are chosen for optimization in multi-user communication systems, they are all used to improve the performance of the communication function.  While for the proposed integrated system, we select the appropriate performance metrics based on the functions of sensing, computing and communication. By solving the formulated multi-objective optimization problem (MOOP), the desired performance of three different functions can be effectively achieved.

\section{Joint Design of Sensing, Computing and Communication}
This section aims at jointly designing transmit and receive beamforming vectors for ISCC in 6G wireless networks. Since 6G wireless networks have different priorities among sensing, computing and communication for various application scenarios, we formulate two typical categories of MOOPs to investigate the trade-offs among them. The first one is a weighted overall performance maximization (WOPM) subject to the maximum budget of transmit power. The second one is a total transmit power minimization (TTPM) while guaranteeing the quality of service (QoS) requirements on sensing, computing, and communication, respectively.

\subsection{Weighted Overall Performance Maximization Design}
Now, we first study three individual single-objective optimization problem (SOOP) for sensing, computing, and communication in 6G wireless networks, respectively, which severs as building blocks for the formulation of WOPM design.  The first SOOP aims at minimizing the weighted sum-MSE for sensing subject to the maximum transmit power budget, which is formulated as
\emph{S-1:  Weighted Sum-MSE of Sensing Minimization:}
\begin{eqnarray}\label{OP1}
\!\!\!\!\!\!\underset{\begin{smallmatrix}
 {{\mathbf{a}}_{k,i}},{{\mathbf{b}}_{k,l}}, \\
 {{\mathbf{c}}_{k,j}},{{\mathbf{v}}_{i}}
\end{smallmatrix}}{\mathop{\text{min }}}\,\!\!\!\!\!\!&&\!\!\!\!\!\sum\limits_{i=1}^{I}{\theta _{i}^{\text{sens}}\text{MSE}_{i}^{\text{sens}}}\\
\!\!\!\textrm{s.t.}\!\!\!\!\!\!&&\!\!\!\!\!\text{C1:}\sum\limits_{i=1}^{I}{{{\left\| {{\mathbf{a}}_{k,i}} \right\|}^{2}}}+\sum\limits_{l=1}^{L}{{{\left\| {{\mathbf{b}}_{k,l}} \right\|}^{2}}}+\sum\limits_{j=1}^{J}{{{\left\| {{\mathbf{c}}_{k,j}} \right\|}^{2}}}\le {{P}_{\max ,k}},\nonumber
\end{eqnarray}
where constant $\theta_{i}^{\text{sens}}$ is the weighted coefficient related to $\text{MSE}_{i}^{\text{sens}}$, and $P_{\max,k}$ is the maximum transmit power budget at the $k$-th sensor. The objective in \emph{S-1} is to minimize the weighted summation of MSEs between estimated reflection coefficients and actual ones for all $I$ targets depicted in (\ref{MSE_sens}), and constraint C1 is the transmit power limitation at the $k$-th sensor.  Note that each SOOP only concerns one performance metric of the proposed integrated sensing, computing and communication without considering others. For example, it is seen from \emph{S-1} that the communication beam ${{\mathbf{c}}_{k,j}}$ and the computing beam ${{\mathbf{b}}_{k,l}}$ will naturally be zero to minimize the objective, while the sensing beam ${{\mathbf{a}}_{k,i}}$ will not be zero due to the term ${{\left| \sum\limits_{k=1}^{K}{\mathbf{v}_{i}^{H}{{\mathbf{G}}_{k,i}}{{\mathbf{a}}_{k,i}}}-1 \right|}^{2}}R_{\text{i}}^{2}$ in (\ref{MSE_sens}). Hence, no additional conditions related to communication and computing beams are required. The second SOOP is to minimize the weighted sum-MSE for computing under the limitation of transmit power consumption, which can be mathematically expressed as

\emph{S-2: Weighted Sum-MSE of Computing Minimization:}
\begin{eqnarray}\label{OP2}
\underset{\begin{smallmatrix}
 {{\mathbf{a}}_{k,i}},{{\mathbf{b}}_{k,l}}, \\
 {{\mathbf{c}}_{k,j}},{{\mathbf{z}}_{l}}
\end{smallmatrix}}{\mathop{\text{min }}}\,\!\!\!\!&&\!\!\!\!\sum\limits_{l=1}^{L}{\theta _{l}^{\text{comp}}\text{MSE}_{l}^{\text{comp}}} \nonumber\\
\textrm{s.t.}&&\!\!\!\!\!\text{C1},
\end{eqnarray}
where constant $\theta_{l}^{\text{comp}}$ is the weighted coefficient relevant to $\text{MSE}_{l}^{\text{comp}}$, and the objective is minimizing the weighted summation of the computation distortions for $L$ model parameters, which was defined in (\ref{MSE_comp}). The third SOOP focuses on the weighted sum-rate maximization for communication subject to the transmit power limitation, which is formulated as

\emph{S-3': Weighted Sum-Rate of Communication Maximization:}
\begin{eqnarray}\label{OP3}
\underset{\begin{smallmatrix}
 {{\mathbf{a}}_{k,i}},{{\mathbf{b}}_{k,l}}, \\
 {{\mathbf{c}}_{k,j}},{{\mathbf{u}}_{k,j}}
\end{smallmatrix}}{\mathop{\text{max}}}\,\!\!\!\!&&\!\!\!\! \sum\limits_{k=1}^{K}{\sum\limits_{j=1}^{J}{\theta _{k,j}^{\text{comm}}{{r}_{k,j}}}} \nonumber\\
\textrm{s.t.}&&\!\!\!\!\!\text{C1},
\end{eqnarray}
where ${{r}_{k,j}}={{\log }_{2}}\left( 1+{{\Gamma }_{k,j}} \right)$ is the achievable rate of the $j$-th communication signal from the $k$-th sensor and $\theta_{k,j}^{\text{comm}}$ is the weighted coefficient associated with  ${{r}_{k,j}}$. \emph{S-3'} is a classical weighted sum-rate maximization problem which is non-convex due to the complicated objective function involving log functions of fractions \cite{NP2}. Moreover, the transmit beamforming vectors $\{{{\mathbf{a}}_{k,i}},{{\mathbf{b}}_{k,l}},{{\mathbf{c}}_{k,j}}\}$ and communication receive beamforming vector $\mathbf{u}_{k,j}$ are coupled in the constraints and objective function.  To handle this issue as well as facilitate the design, we transform \emph{S-3'} into its equivalent log-MSE minimization problem via the following theorem.

\emph{Theorem 1:} The relationship between the received SINR $\Gamma_{k,j}$ and the minimum MSE (MMSE) $e_{k,j}^{\text{comm}}$ for the communication signal $s_{k,j}^{\text{comm}}$ can be expressed as
\begin{equation}\label{Theorem_1}
1+{{\Gamma }_{k,j}}= ({e_{k,j}^{\text{comm}}})^{-1}, \forall k,j.
\end{equation}
\begin{IEEEproof}
 Please refer to Appendix A.
\end{IEEEproof}
According to Theorem 1, we can transform the objective function of \emph{S-3'} to
\begin{equation}\label{tran_obj}
  \underset{\begin{smallmatrix}
 {{\mathbf{a}}_{k,i}},{{\mathbf{b}}_{k,l}}, \\
 {{\mathbf{c}}_{k,j}},{{\mathbf{u}}_{k,j}}
\end{smallmatrix}}{\mathop{\text{min}}}\,\sum\limits_{k=1}^{K}{\sum\limits_{j=1}^{J}{\theta _{k,j}^{\text{comm}}{{\log }_{2}}\left( e_{k,j}^{\text{comm}} \right)}},
\end{equation}
which is equivalent to minimizing the weighted sum-MSE of communication $\text{MSE}_{k,j}^{\text{comm}}$ with the MMSE receiver ${{\mathbf{u}}_{k,j}}=\Xi^{-1}{{\mathbf{H}}_{k}}{{\mathbf{c}}_{k,j}}$, given in Appendix A. Thus, (\ref{tran_obj}) can be reformulated as
 \begin{equation}\label{tran_obj2}
 \underset{{{\mathbf{a}}_{k,i}},{{\mathbf{b}}_{k,l}},{{\mathbf{c}}_{k,j}}}{\text{min}}\, \sum\limits_{k=1}^{K}{\sum\limits_{j=1}^{J}{\theta _{k,j}^{\text{comm}}{{\log }_{2}}\left( \text{MSE}_{k,j}^{\text{comm}} \right)}}
\end{equation}
However, the transformed objective function (\ref{tran_obj2}) still remains non-convex caused by the structure of sum of logarithmic function. To this end, we introduce a weight variable $\omega_{k,j}$ for the $\text{MSE}_{k,j}^{\text{comm}}$ \cite{WMMSE}, and then (\ref{tran_obj2}) is transformed as
 \begin{equation}\label{tran_obj3}
\underset{\begin{smallmatrix}
 {{\mathbf{a}}_{k,i}},{{\mathbf{b}}_{k,l}}, \\
 {{\mathbf{c}}_{k,j,}},{{\omega }_{k,j}}
\end{smallmatrix}}{\mathop{\min }}\,\sum\limits_{k=1}^{K}{\sum\limits_{j=1}^{J}{\theta _{k,j}^{\text{comm}}\left[ {{\omega }_{k,j}}\text{MSE}_{k,j}^{\text{comm}}-{{\log }_{2}}\left( {{\omega }_{k,j}} \right) \right]}}.
\end{equation}
As for unconstrained optimization problems in terms of the weight variable $\omega_{k,j}$, by letting the first-derivative of the objective function (\ref{tran_obj3}) equal to zero, we can obtain the optimal solution as follows
\begin{equation}\label{weight}
  \omega _{k,j}^{*}={{\left( \ln 2\cdot \text{MSE}_{k,j}^{\text{comm}} \right)}^{-1}},
\end{equation}
which makes (\ref{tran_obj2}) and (\ref{tran_obj3}) are equivalent. After the above conversion, \emph{S-3'} can be transformed as

\emph{S-3: Weighted Sum-Modified MSE of Communication Minimization:}
\begin{eqnarray}\label{OPM3}
\underset{\begin{smallmatrix}
 {{\mathbf{a}}_{k,i}},{{\mathbf{b}}_{k,l}},{{\mathbf{c}}_{k,j}}, \\
 {{\mathbf{u}}_{k,j}},{{\omega }_{k,j}}
\end{smallmatrix}}{\mathop{\text{min }}}\,\!\!\!\!&&\!\!\!\!\sum\limits_{k=1}^{K}{\sum\limits_{j=1}^{J}{\theta _{k,j}^{\text{comm}}\left[ {{\omega }_{k,j}}\text{MSE}_{k,j}^{\text{comm}}-{{\log }_{2}}\left( {{\omega }_{k,j}} \right) \right]}} \nonumber\\
\textrm{s.t.}&&\!\!\!\! \text{C1}.
\end{eqnarray}
 To jointly optimize the three performance of sensing, computing, and communication, we adopted the weighted sum method to formulate the corresponding MOOP \cite{MOOP,Ata_WCNC}. The WOPM is formulated as follows:

\emph{M-1: Weighted Overall Performance Maximization:}
\begin{eqnarray}\label{OP4}
\underset{\begin{smallmatrix}
 {{\mathbf{a}}_{k,i}},{{\mathbf{b}}_{k,l}},{{\mathbf{c}}_{k,j}}, \\
 {{\mathbf{v}}_{i}},{{\mathbf{u}}_{k,j}},{{\mathbf{z}}_{l}},{{\omega }_{k,j}}
\end{smallmatrix}}{\mathop{\min }}\,\!\!\!\!&&\!\!\!\!{{\alpha }_{1}}{{\Psi }_{1}}+{{\alpha }_{2}}{{\Psi }_{2}}+{{\alpha }_{3}}{{\Psi }_{3}}\nonumber\\
\textrm{s.t.}&&\!\!\!\! \text{C1},
\end{eqnarray}
where ${{\Psi }_{p}}, p=1,2,3$ is the normalized objective function\footnote{Since the objective value of \emph{S-1}, \emph{S-2} and \emph{S-3} have different ranges, it is desired to perform the normalization for each objective function in order to coordinate the performance of sensing, computing, and communication as well as to facilitate the convergence of objective value for \emph{M-1}. In this paper, we define $\mathcal{F}_p, p=1,2,3,$ as the objective function of \emph{S-$p$} and ${{\Psi }_{p}}={\left( {\mathcal{F}_{p}}-\mathcal{F}_{p}^{*} \right)}/{\left| \mathcal{F}_{p}^{*} \right|}\;$, where $\mathcal{F}_{p}^{*}$ are the corresponding performance limits of sensing, computing and communication in \emph{M-1}, which can be obtained by respectively solving \emph{S-$p$} by applying the same algorithm for solving \emph{M-1}.} of \emph{S-1}, \emph{S-2} and \emph{S-3}, respectively. $\alpha_{p}\geq0 $ is the priority of the $p$-th objective function, which represents the preference of the system operator and is satisfied with $\alpha_{1}+\alpha_{2}+\alpha_{3}=1$. By varying the value of $\alpha_{p}$, \emph{M-1} can yield different solutions. Note that \emph{M-1} is equivalent to \emph{S-p} when $\alpha_{p}=1$ and $ \alpha_q=0, \forall p\neq q$, which means \emph{M-1} is a general formulation of \emph{S-1}, \emph{S-2} and \emph{S-3}. Hence, we aim at solving \emph{M-1} with given priorities $\alpha_p$ in the following.

 Since multiple variables are inter-coupled in the objective function of \emph{M-1}, i.e., transmit beams $\{\mathbf{a}_{k,i},{{\mathbf{b}}_{k,l}},{{\mathbf{c}}_{k,j}}\}$, receive beams $\{{{\mathbf{v}}_{i}},\mathbf{u}_{k,j},{{\mathbf{z}}_{l}}\}$, and weight variables $\{\omega_{k,j}\}$, \emph{M-1} is non-convex, which makes it impossible to obtain an optimal solution in polynomial time. In this context, we turn to find a sub-optimal solution for exploring the trade-off relationship among sensing, computing, and communication in 6G wireless networks. By examining \emph{M-1}, although it is not a joint convex function of all variables, it is a convex one in terms of transmit beams, receive beams, and weight variables, respectively. Based on this observation, an alternating optimization (AO) method is applied to divide \emph{M-1} into three subproblems, i.e., optimizing receive beams by fixing transmit beams and weight variables, optimizing weight variables by fixing transmit and receive beams, and optimizing transmit beams by fixing receive beams and weight variables.  In particular, enabled with the AO method, solution procedure for \emph{M-1} will stop until the objective value converges in the iterations  \cite{rate1}. Let us first address the subproblem of optimizing receive beams $\{{{\mathbf{v}}_{i}},\mathbf{u}_{k,j},{{\mathbf{z}}_{l}}\}$ while other variables remain fixed. By applying the Karush-Kuhn-Tucher (KKT) conditions in \emph{M-1}, i.e.,
 \begin{eqnarray}\label{first_order}
\begin{matrix}
   \frac{\partial }{\partial {{\mathbf{v}}_{i}}}\left( \sum\limits_{p=1}^{3}{{{\alpha }_{p}}{{\Psi }_{p}}} \right)=\mathbf{0}\Rightarrow \frac{\partial }{\partial {{\mathbf{v}}_{i}}}\left( \text{MSE}_{i}^{\text{sens}} \right)=\mathbf{0},  \\
   \frac{\partial }{\partial {{\mathbf{z}}_{l}}}\left( \sum\limits_{p=1}^{3}{{{\alpha }_{p}}{{\Psi }_{p}}} \right)=\mathbf{0}\Rightarrow \frac{\partial }{\partial {{\mathbf{z}}_{l}}}\left( \text{MSE}_{l}^{\text{comp}} \right)=\mathbf{0},  \\
   \frac{\partial }{\partial {{\mathbf{u}}_{k,j}}}\left( \sum\limits_{p=1}^{3}{{{\alpha }_{p}}{{\Psi }_{p}}} \right)=\mathbf{0}\Rightarrow \frac{\partial }{\partial {{\mathbf{u}}_{k,j}}}\left( \text{MSE}_{k,j}^{\text{comm}} \right)=\mathbf{0},  \\
\end{matrix}
 \end{eqnarray}
  we can acquire the optimal receive beamforming vectors, also called MMSE receiver, as below
\begin{equation}\label{receiverV}
{{\mathbf{v}}_{i}}={{\left[ {{\bm{\Phi }}_{i}}\left( \sum\limits_{k=1}^{K}{\mathbf{a}_{k,i}^{H}\mathbf{G}_{k,i}^{H}} \right)+\bm{\Phi } \right]}^{-\text{1}}}{{\bm{\Phi }}_{i}},
\end{equation}
\begin{equation}\label{receiverZ}
{{\mathbf{z}}_{l}}\text{=}{\bm{\Xi }^{-1}}\sum\limits_{k=1}^{K}{{{\mathbf{H}}_{k}}{{\mathbf{b}}_{k,l}}},
\end{equation}
and
\begin{equation}\label{receiverU}
{{\mathbf{u}}_{k,j}}={\bm{\Xi }^{-1}}{{\mathbf{H}}_{k}}{{\mathbf{c}}_{k,j}},
\end{equation}
respectively, where $\bm{\Phi} =\sum\limits_{k=1}^{K}{\sum\limits_{n=1,n\ne i}^{I}{R_{n}^{2}{{\mathbf{G}}_{k,n}}{{\mathbf{a}}_{k,n}}\mathbf{a}_{k,n}^{H}\mathbf{G}_{k,n}^{H}}}+\sum\limits_{k=1}^{K}{\sum\limits_{o=1}^{O}{\sum\limits_{m=1}^{I}{R_{o}^{2}{{\mathbf{F}}_{k,o}}{{\mathbf{a}}_{k,m}}\mathbf{a}_{k,m}^{H}\mathbf{F}_{k,o}^{H}}}}+\sum\limits_{k=1}^{K}{\sum\limits_{l=1}^{L}{{{\mathbf{H}}_{k}}{{\mathbf{b}}_{k,l}}\mathbf{b}_{k,l}^{H}\mathbf{H}_{k}^{H}}}+\sum\limits_{k=1}^{K}{\sum\limits_{m=1}^{J}{{{\mathbf{H}}_{k}}{{\mathbf{c}}_{k,m}}}}\mathbf{c}_{k,m}^{H}\mathbf{H}_{k}^{H}+\sigma _{n}^{2}{{\mathbf{I}}_{N}}$, ${{\bm{\Phi }}_{i}}=R_{i}^{2}\sum\limits_{k=1}^{K}{{{\mathbf{G}}_{k,i}}{{\mathbf{a}}_{k,i}}}$, and $\bm{\Xi} =\bm{\Phi} +\sum\limits_{k=1}^{K}{\sum\limits_{n=1}^{I}{R_{n}^{2}{{\mathbf{G}}_{k,n}}{{\mathbf{a}}_{k,n}}\mathbf{a}_{k,n}^{H}\mathbf{G}_{k,n}^{H}}}$ is defined in Appendix A.  Next, for the second subproblem in terms of optimizing weight variables $\{\omega_{k,j}\}$, the optimal solution is given in (\ref{weight}), i.e., $\omega _{k,j}^{*}={{\left( \ln 2\cdot \text{MSE}_{k,j}^{\text{comm}} \right)}^{-1}}$. Finally, with the MMSE receivers and weight variables, the last subproblem of optimizing transmit beams $\{\mathbf{a}_{k,i},{{\mathbf{b}}_{k,l}},{{\mathbf{c}}_{k,j}}\}$ for maximizing the weighted overall performance is a standard convex quadratic constrained quadratic programming (QCQP) problem, which can be solved by an inter-point method (IPM) [\cite{Convex}, Chapter 11]. At first, we utilize the barrier method to transform this subproblem into an unconstrained convex optimization problem by adding the logarithmic barrier function. Then, the Newton’s method can be applied to obtain the solution. Specifically, the transformed unconstrained convex optimization problem can be written as
\begin{equation}\label{OPnew}
\underset{{{\mathbf{a}}_{k,i}},{{\mathbf{b}}_{k,l}},{{\mathbf{c}}_{k,j}}}{\mathop{\min }}\,\varepsilon(\alpha_1{{\Psi }_{1}}+\alpha_2{{\Psi }_{2}}+\alpha_3{{\Psi }_{3}})-\sum\limits_{k=1}^{K}{\log \left( -f_{k}^{\text{con}} \right)},
\end{equation}
where $f_{k}^{\text{con}}=\sum\limits_{i=1}^{I}{{{\left\| {{\mathbf{a}}_{k,i}} \right\|}^{2}}}+\sum\limits_{l=1}^{L}{{{\left\| {{\mathbf{b}}_{k,l}} \right\|}^{2}}}+\sum\limits_{j=1}^{J}{{{\left\| {{\mathbf{c}}_{k,j}} \right\|}^{2}}}-{{P}_{\max ,k}}$ and $\varepsilon>0$ is a barrier parameter that sets the accuracy of the approximation. Then, we compute the variables' gradients and Hessian for Newton’s method as follows
\begin{eqnarray}
\begin{matrix}
   {{\nabla }_{{{\mathbf{a}}_{k,i}}}}=2\varepsilon\left( \mathbf{T}_{k,i}^{a}{{\mathbf{a}}_{k,i}}-R_{\text{i}}^{2}{{{\tilde{\alpha }}}_{1}}\theta _{i}^{\text{sens}}\mathbf{G}_{k,i}^{H}{{\mathbf{v}}_{i}} \right)+\frac{2{{\mathbf{a}}_{k,i}}}{f_{k}^{\text{con}}},  \\
   {{\nabla }_{{{\mathbf{b}}_{k,l}}}}=\frac{2\varepsilon}{{{K}^{2}}}\left( \mathbf{T}_{k,l}^{{b}}{{\mathbf{b}}_{k,l}}-{{{\tilde{\alpha }}}_{2}}\theta _{l}^{\text{comp}}\mathbf{H}_{k}^{H}{{\mathbf{z}}_{l}} \right)+\frac{2{{\mathbf{b}}_{k,l}}}{f_{k}^{\text{con}}},  \\
   {{\nabla }_{{{\mathbf{c}}_{k,j}}}}=2\varepsilon\left( \mathbf{T}_{k,l}^{{c}}{{\mathbf{c}}_{k,j}}-{{{\tilde{\alpha }}}_{3}}\theta _{k,j}^{\text{comm}}{{\omega }_{k,j}}\mathbf{H}_{k}^{H}{{\mathbf{u}}_{k,j}} \right)+\frac{2{{\mathbf{c}}_{k,j}}}{f_{k}^{\text{con}}},  \\
\end{matrix}
\end{eqnarray}
and
\begin{eqnarray}
\begin{matrix}
   \nabla _{{{\mathbf{a}}_{k,i}}}^{2}=2\varepsilon\mathbf{T}_{k,i}^{{a}}+\frac{4}{{{\left( f_{k}^{\text{con}} \right)}^{2}}}{{\mathbf{a}}_{k,i}}\mathbf{a}_{k,l}^{T}+\frac{2}{f_{k}^{\text{con}}}\mathbf{I},  \\
   \nabla _{{{\mathbf{b}}_{k,l}}}^{2}=\frac{2\varepsilon}{{{K}^{2}}}\mathbf{T}_{k,l}^{{b}}+\frac{4}{{{\left( f_{k}^{\text{con}} \right)}^{2}}}{{\mathbf{b}}_{k,l}}\mathbf{b}_{k,l}^{T}+\frac{2}{f_{k}^{\text{con}}}\mathbf{I},  \\
   \nabla _{{{\mathbf{c}}_{k,j}}}^{2}=2\varepsilon\mathbf{T}_{k,l}^{{c}}+\frac{4}{{{\left( f_{k}^{\text{con}} \right)}^{2}}}{{\mathbf{c}}_{k,j}}\mathbf{c}_{k,j}^{T}+\frac{2}{f_{k}^{\text{con}}}\mathbf{I},  \\
\end{matrix}
\end{eqnarray}
with
\begin{eqnarray}\label{Delta_a}
\!\!\!\!\!\!\mathbf{T} _{k,i}^{{a}}\!\!\!&=&\!\!\!{{{\tilde{\alpha }}}_{1}}\theta _{i}^{\text{sens}}\left( R_{\text{i}}^{2}\mathbf{G}_{k,i}^{H}{{\mathbf{v}}_{i}}\mathbf{v}_{i}^{H}{{\mathbf{G}}_{k,i}}+\sum\limits_{o=1}^{O}{R_{o}^{2}}\mathbf{F}_{k,o}^{H}{{\mathbf{v}}_{i}}\mathbf{v}_{i}^{H}{{\mathbf{F}}_{k,o}} \right)\nonumber\\
\!\!\!&+&\!\!\!\frac{{{{\tilde{\alpha }}}_{2}}}{{{K}^{2}}}\sum\limits_{l=1}^{L}{\theta _{l}^{\text{comp}}\left( R_{i}^{2}\mathbf{G}_{k,i}^{H}{{\mathbf{z}}_{l}}\mathbf{z}_{l}^{H}{{\mathbf{G}}_{k,i}}+\sum\limits_{o=1}^{O}{R_{o}^{2}}\mathbf{F}_{k,o}^{H}{{\mathbf{z}}_{l}}\mathbf{z}_{l}^{H}{{\mathbf{F}}_{k,o}} \right)}\nonumber \\
\!\!\!&+&\!\!\!{{{\tilde{\alpha }}}_{3}}\sum\limits_{k=1}^{K}{\sum\limits_{j=1}^{J}{\theta _{k,j}^{\text{comm}}{{\omega }_{k,j}} R_{i}^{2}\mathbf{H}_{k}^{H}{{\mathbf{u}}_{k,j}}\mathbf{u}_{k,j}^{H}{{\mathbf{H}}_{k}}}},\nonumber\\
\!\!\!&+&\!\!\!{{{\tilde{\alpha }}}_{3}}\sum\limits_{k=1}^{K}{\sum\limits_{j=1}^{J}{\theta _{k,j}^{\text{comm}}{{\omega }_{k,j}} \sum\limits_{o=1}^{O}{R_{o}^{2}}\mathbf{F}_{k,o}^{H}{{\mathbf{u}}_{k,j}}\mathbf{u}_{k,j}^{H}{{\mathbf{F}}_{k,o}} }},
\end{eqnarray}
\begin{eqnarray}\label{Delta_b}
\mathbf{T_{k,l}^{{b}}}\!\!\!&=&\!\!\!{{{\tilde{\alpha }}}_{1}}{{K}^{2}}\sum\limits_{i=1}^{I}{\theta _{i}^{\text{sens}}\mathbf{H}_{k}^{H}{{\mathbf{v}}_{i}}\mathbf{v}_{i}^{H}{{\mathbf{H}}_{k}}}+{{{\tilde{\alpha }}}_{2}}\theta _{l}^{\text{comp}}\mathbf{H}_{k}^{H}{{\mathbf{z}}_{l}}\mathbf{z}_{l}^{H}{{\mathbf{H}}_{k}}\nonumber\\
\!\!\!&+&\!\!\!{{\tilde{\alpha }}}_{3}{{K}^{2}}\sum\limits_{k=1}^{K}{\sum\limits_{j=1}^{J}{\theta _{k,j}^{\text{comm}}{{\omega }_{k,j}}\mathbf{H}_{k}^{H}{{\mathbf{u}}_{k,j}}\mathbf{u}_{k,j}^{H}{{\mathbf{H}}_{k}}}},
\end{eqnarray}
and
\begin{eqnarray}\label{Delta_c}
\mathbf{T} _{k,j}^{{c}}\!\!\!&=&\!\!\!{{{\tilde{\alpha }}}_{1}}\sum\limits_{i=1}^{I}{\theta _{i}^{\text{sens}}\mathbf{H}_{k}^{H}{{\mathbf{v}}_{i}}\mathbf{v}_{i}^{H}\mathbf{H}}+\frac{{{{\tilde{\alpha }}}_{2}}}{{{K}^{2}}}\sum\limits_{l=1}^{L}{\theta _{l}^{\text{comp}}\mathbf{H}_{k}^{H}{{\mathbf{z}}_{l}}\mathbf{z}_{l}^{H}{{\mathbf{H}}_{k}}}\nonumber\\
\!\!\!&+&\!\!\!{{{\tilde{\alpha }}}_{3}}\theta _{k,j}^{\text{comm}}{{\omega }_{k,j}}\mathbf{H}_{k}^{H}{{\mathbf{u}}_{k,j}}\mathbf{u}_{k,j}^{H}{{\mathbf{H}}_{k}},
\end{eqnarray}
where $\nabla_\mathbf{x}$ and $\nabla_\mathbf{x}^{2}$ denote the gradient and the Hessian of the variable for the objective function (\ref{OPnew}), respectively. Moreover, ${{{\tilde{\alpha}}}_{1}}=\frac{{{\alpha }_{1}}}{\left| \mathcal{F}_{1}^{*} \right|}$, ${{{\tilde{\alpha }}}_{1}}=\frac{{{\alpha }_{2}}}{\left| \mathcal{F}_{2}^{*} \right|}$, and ${{{\tilde{\alpha }}}_{3}}=\frac{{{\alpha }_{3}}}{\left| \mathcal{F}_{3}^{*} \right|}$. With the gradients and Hessian, the transmit beams can be computed by the following updating rules:
\begin{eqnarray}
 \mathbf{a}_{k,i}^{(t+1)}&=&\mathbf{a}_{k,i}^{(t)}-\eta _{k,i}^{a}{{\left( \nabla _{{{\mathbf{a}}_{k,i}}}^{2} \right)}^{-1}}{{\nabla }_{{{\mathbf{a}}_{k,i}}}} \label{beam_a}\\
  \mathbf{b}_{k,l}^{(t+1)}&=&\mathbf{b}_{k,l}^{(t)}-\eta _{k,l}^{b}{{\left( \nabla _{{{\mathbf{b}}_{k,l}}}^{2} \right)}^{-1}}{{\nabla }_{{{\mathbf{b}}_{k,l}}}} \label{beam_b}\\
 \mathbf{c}_{k,j}^{(t+1)}&=&\mathbf{c}_{k,l}^{(t)}-\eta _{k,l}^{c}{{\left( \nabla _{{{\mathbf{c}}_{k,j}}}^{2} \right)}^{-1}}{{\nabla }_{{{\mathbf{c}}_{k,j}}}} \label{beam_c}
\end{eqnarray}
where $\eta _{k,i}^{a},\eta _{k,i}^{b}$ and $\eta _{k,i}^{c}$ are the step sizes of the transmit beams $\mathbf{a}_{k,i}, \mathbf{b}_{k,l}$ and $\mathbf{c}_{k,j}$, respectively. In summary, the joint design of ISCC for 6G wireless networks based on the WOPM is concluded as Algorithm \ref{alg1}.
\begin{algorithm}[t]
\small
\caption{: WOPM-based Joint Design of ISCC for 6G wireless networks}
\label{alg1}
\hspace*{0.02in} {\bf Input:} %算法的结果输入 \forall k,i,l,j,p
 $N,M,K,O,I,L,J,\sigma_n^2,{{\theta }_{k,i}^{\text{sens}}},{{\theta }_{k,l}^{\text{comp}}},{{\theta }_{k,j}^{\text{comm}}},P_{\max,k},\alpha_p$\\
\hspace*{0.02in} {\bf Output:} %算法的结果输出
${{\mathbf{a}}_{k,i}},{{\mathbf{b}}_{k,l}},{{\mathbf{c}}_{k,j}},{{\mathbf{v}}_{i}},\mathbf{u}_{k,j},{{\mathbf{z}}_{l}} $
\begin{algorithmic}[1]
\STATE{\textbf{Initialize} Iteration index $t=1$, $\mathbf{a}_{k,i}^{(0)}=\mathbf{b}_{k,l}^{(0)}=\mathbf{c}_{k,j}^{(0)}=[\sqrt{\frac{P_{\max,k}}{3M}},0,\ldots,0]^{T},\forall k,i,l,j$;}
\REPEAT
 \STATE{Compute $\mathbf{v}_{i}^{(t)}$ according to (\ref{receiverV}) with $\mathbf{a}_{k,i}^{(t-1)}, \mathbf{b}_{k,l}^{(t-1)}$ and $\mathbf{c}_{k,j}^{(t-1)}$;}
 \STATE{Compute $\mathbf{z}_{l}^{(t)}$ according to (\ref{receiverZ}) with $\mathbf{a}_{k,i}^{(t-1)}, \mathbf{b}_{k,l}^{(t-1)}$ and $\mathbf{c}_{k,j}^{(t-1)}$;}
 \STATE{Compute $\mathbf{u}_{k,j}^{(t)}$ according to (\ref{receiverU}) with $\mathbf{a}_{k,i}^{(t-1)}, \mathbf{b}_{k,l}^{(t-1)}$ and $\mathbf{c}_{k,j}^{(t-1)}$;}
 \STATE{Compute $\omega _{k,j}^{(t)}$ according to (\ref{weight}) with $\mathbf{v}_{i}^{(t)}$, $\mathbf{z}_{l}^{(t)}$, $\mathbf{u}_{k,j}^{(t)}$, $\mathbf{a}_{k,i}^{(t-1)}, \mathbf{b}_{k,l}^{(t-1)}$ and $\mathbf{c}_{k,j}^{(t-1)}$ ;}
 \STATE{\textbf{Initialize} Iteration index $m=1$, barrier parameter $\varepsilon=1$, amplification factor $\nu$=10;}
\REPEAT
 \STATE{Select feasible step sizes $\eta_{k,l}^{a(m)}, \eta_{k,l}^{b(m)}$ and $\eta_{k,l}^{c(m)}$;}
 \STATE{Compute $\mathbf{a}_{k,i}^{(m)}$ according to (\ref{beam_a}) with $\eta_{k,l}^{a(m)}$,$\mathbf{a}_{k,i}^{(m-1)}$,$\mathbf{v}_{k,i}^{(t)}, \mathbf{z}_{l}^{(t)},\mathbf{u}_{k,j}^{(t)}$ and $\omega _{k,j}^{(t)}$;}
\STATE{Compute $\mathbf{b}_{k,l}^{(m)}$ according to (\ref{beam_b}) with $\eta_{k,l}^{b(m)}$,$\mathbf{b}_{k,l}^{(m-1)}$,$\mathbf{v}_{k,i}^{(t)}, \mathbf{z}_{l}^{(t)},\mathbf{u}_{k,j}^{(t)}$ and $\omega _{k,j}^{(t)}$;}
\STATE{Compute $\mathbf{c}_{k,j}^{(m)}$ according to (\ref{beam_c}) with $\eta_{k,l}^{c(m)}$,$\mathbf{c}_{k,j}^{(m-1)}$,$\mathbf{v}_{k,i}^{(t)}, \mathbf{z}_{l}^{(t)},\mathbf{u}_{k,j}^{(t)}$ and $\omega _{k,j}^{(t)}$;}
 \IF{Meeting the centrality condition}
 \STATE{Update $\varepsilon=\varepsilon*\nu$;}
\ENDIF
   \STATE{Update $m=m+1$;}
\UNTIL{The duality gap converges}
\STATE{Update $\mathbf{a}_{k,i}^{(t)}=\mathbf{a}_{k,i}^{(m)},\mathbf{b}_{k,l}^{(t)}=\mathbf{b}_{k,l}^{(m)}$ and $\mathbf{c}_{k,j}^{(t)}=\mathbf{c}_{k,j}^{(m)}$;}
 \STATE{Update $t=t+1$;}
\UNTIL{The objective value converges}
 \end{algorithmic}
\end{algorithm}

%Proposed_Solution
\subsection{Total Transmit Power Minimization Design}
Now, let us focus on another MOOP with the goal of minimizing total transmit power consumption but ensuring the QoS requirements on sensing, computing, and communication, respectively. The design is formulated as follows:

\emph{M-2: Total Transmit Power Minimization:}
\begin{eqnarray}\label{OP5}
\underset{\begin{smallmatrix}
 {{\mathbf{a}}_{k,i}},{{\mathbf{b}}_{k,l}},{{\mathbf{c}}_{k,j}} \\
 ,{{\mathbf{v}}_{i}},{{\mathbf{u}}_{k,j}},{{\mathbf{z}}_{l}}
\end{smallmatrix}}{\mathop{\text{min }}}\,\!\!\!\!&&\!\!\!\! \sum\limits_{k=1}^{K}{\left( \sum\limits_{i=1}^{I}{{{\left\| {{\mathbf{a}}_{k,i}} \right\|}^{2}}}+\sum\limits_{l=1}^{L}{{{\left\| {{\mathbf{b}}_{k,l}} \right\|}^{2}}}+\sum\limits_{j=1}^{J}{{{\left\| {{\mathbf{c}}_{k,j}} \right\|}^{2}}} \right)} \nonumber\\
\textrm{s.t.}&&\!\!\!\! \text{C2:\ } \text{ MSE}_{i}^{\text{sens}}\le {{\delta }_{i}},\forall i \in \Omega_I,\nonumber\\
&&\!\!\!\!\text{C3: \ } \text{MSE}_{l}^{\text{comp}}\le {{\chi }_{l}},\forall l \in \Omega_L,\nonumber\\
&&\!\!\!\!\text{C4: \ } {{\Gamma }_{k,j}}\ge {{\gamma }_{k,j}},\forall k \in \Omega_K, j \in \Omega_J,
\end{eqnarray}
where  ${{\delta }_{i}}>0$ is the given tolerable maximum sensing distortion for $i$-th target object, ${{\chi }_{l}}>0$ is the given maximum tolerable computation error for the $l$-th model parameter, and ${{\gamma }_{k,j}}>0$ denotes the required minimum SINR for the communication signal $s_{k,j}^{\text{comm}}$. The objective function of \emph{M-2} is minimizing the total transmit power of all sensors, and the constraints C2, C3, and C4 are the QoS requirements for sensing, computing, and communication, respectively. However, \emph{M-2} is not convex because of the variables coupling in constraints C2, C3 and C4, i.e., transmit beams $\{{{\mathbf{a}}_{k,i}},{{\mathbf{b}}_{k,l}},{{\mathbf{c}}_{k,j}}\}$ and receive beams $\{{{\mathbf{v}}_{i}},\mathbf{u}_{k,j},{{\mathbf{z}}_{l}}\}$, which makes it hard to find the optimal solution in the polynomial time. Thus, it is also desired to divide \emph{M-2} into two subproblems and then solve it by the AO method.  Considering the balance between the system performance and the computational complexity, it is feasible to apply the MMSE receivers in the subproblem for optimizing receive beams by fixing transmit beams, which are given in (\ref{receiverV}), (\ref{receiverZ}), and (\ref{receiverU}). For another subproblem for optimizing transmit beams by fixing receive beams, we need to handle the non-convex constraint C4, which is equivalent to
\begin{eqnarray}\label{SINR0}
\!\!\!\!\!\!\!\!\!\!\!\!\!\!\!&&\frac{1}{{{\gamma }_{k,j}}}{{\left| \mathbf{u}_{k,j}^{H}{{\mathbf{H}}_{k}}{{\mathbf{c}}_{k,j}} \right|}^{2}}\geq \nonumber\\
\!\!\!\!\!\!\!\!\!\!\!\!&&{\sum\limits_{i=1,i\ne k}^{K}{\sum\limits_{n=1,n\ne j}^{J}{{{\left| \mathbf{u}_{k,j}^{H}{{\mathbf{H}}_{i}}{{\mathbf{c}}_{i,n}} \right|}^{2}}}}+{{X}_{k,j}}+\sigma _{n}^{2}{{\left\| \mathbf{u}_{k,j}^{{}} \right\|}^{2}}}.
\end{eqnarray}
To address the non-convexity, we introduce an auxiliary variable ${{\mathbf{C}}_{k,j}}={{\mathbf{c}}_{k,j}}\mathbf{c}_{k,j}^{H}$ and substitute it into (\ref{SINR0}), which is given by
\begin{eqnarray}\label{SINR1}
  \!\!\!\!\!\!\!\!\!\!\!\!\!\!\!&&\frac{1}{{{\gamma }_{k,j}}}\text{tr}\left( \mathbf{u}_{k,j}^{H}{{\mathbf{H}}_{k}}{{\mathbf{C}}_{k,j}}\mathbf{H}_{k}^{H}{{\mathbf{u}}_{k,j}} \right)\geq \\
  \!\!\!\!\!\!\!\!\!\!\!\!&& \sum\limits_{i=1,i\ne k}^{K}{\sum\limits_{n=1,n\ne j}^{J}\text{tr}\left({\mathbf{u}_{k,j}^{H}{{\mathbf{H}}_{i}}{{\mathbf{C}}_{i,n}}\mathbf{H}_{i}^{H}{{\mathbf{u}}_{k,j}}} \right)}+{{X}_{k,j}}+\sigma _{n}^{2}{{\left\| \mathbf{u}_{k,j} \right\|}^{2}}. \nonumber
\end{eqnarray}
Based on this, the subproblem for the transmit beamforming optimization can be converted to a standard semi-definite programming (SDP) problem as follows

\emph{M-2': Equivalent Subproblem of Transmit Beamforming Optimization:}
\begin{eqnarray}\label{OP6}
\underset{{{\mathbf{a}}_{k,i}},{{\mathbf{b}}_{k,l}},{{\mathbf{C}}_{k,j}}}{\mathop{\min }}\,\!\!\!\!&&\!\!\!\!\sum\limits_{k=1}^{K}{\left( \sum\limits_{i=1}^{I}{{{\left\| {{\mathbf{a}}_{k,i}} \right\|}^{2}}}+\sum\limits_{l=1}^{L}{{{\left\| {{\mathbf{b}}_{k,l}} \right\|}^{2}}}+\sum\limits_{j=1}^{J}{\text{tr}\left( {{\mathbf{C}}_{k,j}} \right)} \right)} \nonumber\\
\textrm{s.t.}&&\!\!\!\! \text{C5: \ } \overline{\text{MSE}}_{i}^{\text{sens}}\le {{\delta }_{i}},\forall i \in \Omega_I,\nonumber\\
&&\!\!\!\!\text{C6: \ } \overline{\text{MSE}}_{l}^{\text{comp}}\le {{\chi }_{l}},\forall l \in \Omega_L,\nonumber\\
&&\!\!\!\!\text{C7: \ } (\ref{SINR1}),\nonumber\\
&&\!\!\!\!\text{C8: \ } {{\mathbf{C}}_{k,j}}\succeq \mathbf{0},\forall k \in \Omega_K, j \in \Omega_J,\nonumber\\
&&\!\!\!\!\text{C9: \ } {\text{Rank}\left( {{\mathbf{C}}_{k,j}} \right)=1},\forall k \in \Omega_K, j \in \Omega_J,
\end{eqnarray}
where $\overline{\text{MSE}}_{i}^{\text{sens}}=\text{MSE}_{i}^{\text{sens}}-\sum\limits_{k=1}^{K}{\sum\limits_{j=1}^{J}{{{\left| \mathbf{v}_{i}^{H}{{\mathbf{H}}_{k}}{{\mathbf{c}}_{k,j}} \right|}^{2}}+\sum\limits_{k=1}^{K}{\sum\limits_{j=1}^{J}{\text{tr}\left( \mathbf{v}_{i}^{H}{{\mathbf{H}}_{k}}{{\mathbf{C}}_{k,j}}\mathbf{H}_{k}^{H}{{\mathbf{v}}_{i}} \right)}}}}$ and $\overline{\text{MSE}}_{l}^{\text{comp}}=\text{MSE}_{l}^{\text{comp}}-\frac{1}{K^2}\sum\limits_{k=1}^{K}{\sum\limits_{j=1}^{J}{{{\left| \mathbf{z}_{l}^{H}{{\mathbf{H}}_{k}}{{\mathbf{c}}_{k,j}} \right|}^{2}}+\frac{1}{K^2}\sum\limits_{k=1}^{K}{\sum\limits_{j=1}^{J}{\text{tr}\left( \mathbf{z}_{l}^{H}{{\mathbf{H}}_{k}}{{\mathbf{C}}_{k,j}}\mathbf{H}_{k}^{H}{{\mathbf{z}}_{l}} \right)}}}}$. Note that constraint C9 for the rank-one limitation of ${{\mathbf{C}}_{k,j}}$ is non-convex, which makes \emph{M-2'} still non-convexity. To this end, we adopt the semi-definite relaxation (SDR) technique, namely discarding constraint C9. As a result, \emph{M-2'} is restated as

\emph{M-2'': Transformed Subproblem of Transmit Beamforming Optimization:}
\begin{eqnarray}\label{OP7}
\underset{{{\mathbf{a}}_{k,i}},{{\mathbf{b}}_{k,l}},{{\mathbf{C}}_{k,j}}}{\mathop{\min }}\,\!\!\!\!&&\!\!\!\!\sum\limits_{k=1}^{K}{\left( \sum\limits_{i=1}^{I}{{{\left\| {{\mathbf{a}}_{k,i}} \right\|}^{2}}}+\sum\limits_{l=1}^{L}{{{\left\| {{\mathbf{b}}_{k,l}} \right\|}^{2}}}+\sum\limits_{j=1}^{J}{\text{tr}\left( {{\mathbf{C}}_{k,j}} \right)} \right)} \nonumber\\
\textrm{s.t.}&&\!\!\!\! \text{C5}-\text{C8}.
\end{eqnarray}
It is found that \emph{M-2''} is a joint convex problem in terms of transmit beams $\{{\mathbf{a}_{k,i}},{\mathbf{b}_{k,l}},{\mathbf{C}_{k,j}}\}$ due to the convex constraints C5-C8 and convex objective function, and thus it can be directly solved via CVX \cite{CVX}. It is worth mentioning that for the rank of the optimal solution to \emph{M-2''}, we have following theorem:

\emph{Theorem 2:} The rank of the optimal solution ${{\mathbf{C}}_{k,j}^{*}}$ to \emph{M-2''} always satisfies $\text{Rank}\left( {{\mathbf{C}}_{k,j}^{*}} \right)=1, \forall k,j$.
\begin{IEEEproof}
Please refer to Appendix B.
\end{IEEEproof}
According to Theorem 2, we can perform eigenvalue decomposition (EVD) to get the unique optimal solution ${{\mathbf{c}}_{k,j}^{*}}$ of \emph{M-2}. The EVD operation can be presented as
\begin{equation}\label{EVD}
  {{\mathbf{c}}_{k,j}^{*}}=\sqrt{{{\lambda }_{k,j}^{\max }}({{\mathbf{C}}_{k,j}^{*}})}{{\bm{\xi }}_{k,j}^{\max }},
\end{equation}
where ${{\lambda }_{k,j}^{\max }}({{\mathbf{C}}_{k,j}^{*}})$ denotes the maximum eigenvalue of $\mathbf{C}_{k,j}^{*}$ and ${{\bm{\xi }}_{k,j}^{\max }}$ represents the corresponding unit eigenvector. Hence, the joint design of ISCC for 6G wireless networks based on TTPM is summarized as Algorithm \ref{alg2}.

\begin{algorithm}[th]
\small
\caption{: TTPM-based Joint Design of ISCC for 6G wireless networks}
\label{alg2}
\hspace*{0.02in} {\bf Input:} %,$\forall k,i,l,j$
 $K,N,M,I,O,L,J,\sigma_n^2,P_0,\delta_{i},\chi_{l},\gamma_{k,j}$\\
\hspace*{0.02in} {\bf Output:} %$\forall k,i,l,j$
$\mathbf{a}_{k,i}$, $\mathbf{b}_{k,l}$, $\mathbf{c}_{k,j}$,$\mathbf{v}_i$,$\mathbf{z}_l$,$\mathbf{u}_{k,j}$
\begin{algorithmic}[1]
\STATE{\textbf{Initialize} iteration index $t=1$, $\mathbf{a}_{k,i}^{(0)}= \mathbf{b}_{k,l}^{(0)}=\mathbf{c}_{k,j}^{(0)}=[\sqrt{\frac{P_{0}}{3M}},0,\ldots,0]^{T},\forall k,i,l,j$;}
\REPEAT
 \STATE{Compute $\mathbf{v}_{i}^{(t)}$ according to (\ref{receiverV}) with $\mathbf{a}_{k,i}^{(t-1)}, \mathbf{b}_{k,l}^{(t-1)}$ and $\mathbf{c}_{k,j}^{(t-1)}$;}
 \STATE{Compute $\mathbf{z}_{l}^{(t)}$ according to (\ref{receiverZ}) with $\mathbf{a}_{k,i}^{(t-1)}, \mathbf{b}_{k,l}^{(t-1)}$ and $\mathbf{c}_{k,j}^{(t-1)}$;}
 \STATE{Compute $\mathbf{u}_{k,j}^{(t)}$ according to (\ref{receiverU}) with $\mathbf{a}_{k,i}^{(t-1)}, \mathbf{b}_{k,l}^{(t-1)}$ and $\mathbf{c}_{k,j}^{(t-1)}$;}
 \STATE{Obtain $\{\mathbf{a}^{(t)}_{k,i},\mathbf{b}^{(t)}_{k,l},\mathbf{C}_{k,j}^{(t)}\}$ by solving \emph{M-2''} with fixed $\{\mathbf{v}_{i}^{(t)},\mathbf{z}_l^{(t)},\mathbf{u}_{k,j}^{(t)}\}$; }
 \STATE{Obtain  $\mathbf{c}_{k,j}^{(t)}$ by EVD on $\mathbf{C}_{k,j}^{(t)}$ according to (\ref{EVD});}
 \STATE{$t=t+1$;}
\UNTIL{convergence}
 \end{algorithmic}
\end{algorithm}

\subsection{Convergence and Complexity Analysis of Proposed Algorithms}
Herein, let us discuss the property of convergence and complexity for the proposed two algorithms.

 \emph{Convergence Analysis:} For Algorithm 1, since \emph{M-1} is convex for each optimization variable while fixing the others, the optimal solutions of every subproblem can be found. In fact, the solutions of each iteration are feasible in the next iteration. As a result, the objective value of \emph{M-1} is monotonically non-increasing during the iterations. Furthermore, due to the transmit power constraint at the sensors, the objective value of \emph{M-1} is lower bounded. Hence, based on the monotone bounded convergence (MBC) theorem \cite{Convergence}, Algorithm 1 would converge for a suitable number of iteration. Next, we analyze the convergence property of Algorithm 2.  First, the adopted MMSE receivers $\{ \mathbf{v}_i, \mathbf{z}_l, \mathbf{u}_{k,j} \}$ is effective to ensure the QoS requirements, which can help enhance the system performance. Moreover, since \emph{M-2''} is convex for transmit beams $\{\mathbf{a}_{k,i} \mathbf{b}_{k,l},\mathbf{c}_{k,j}\}$ with given MMSE receivers, it is feasible to get the optimal solution via CVX directly, which ensures the obtained objective value in the $(t+1)$-th iteration is less than that in the $t$-th iteration. That implies the total transmit power consumption is monotonically non-increasing in the iterations. Moreover, due to the QoS requirements in constraint C2, C3, and C4, the total transmit power consumption is lower bounded. Thus, Algorithm 2 would converge according to the MBC theorem.
%It is easily found that Algorithm 2 works as long as the initial values are set suitably, since the optimized objective in Problem 5 is not a function of receive beams.

\emph{Complexity Analysis:} Since Algorithm 1 and Algorithm 2 are both iterative, the execution steps are the same for each iteration. Thus, we focus on analyzing the per-iteration complexity of the proposed algorithms. By noticing the operations of algorithms, it is observed that the dominating computational complexities arise from the steps 7-17 for Algorithm 1, i.e., finding optimal $\{\mathbf{a}^{(t)}_{k,i},\mathbf{b}^{(t)}_{k,l},\mathbf{c}_{k,j}^{(t)}\}$ of \emph{M-1}, and the step 6 for Algorithm 2, i.e., finding optimal $\{\mathbf{a}^{(t)}_{k,i},\mathbf{b}^{(t)}_{k,l},\mathbf{C}_{k,j}^{(t)}\}$ of \emph{M-2''}. Since \emph{M-1} and \emph{M-2''} only contain second-order cone (SOC) or linear matrix inequalities (LMI) constraints, they could both be solved by a standard IPM \cite{Convex}. As such, the worst-case runtime by IPM can be used to depict the computational complexities of algorithm \cite{Lecture_complexity}. Specifically, \emph{M-1} has $K$ SOC constraints of dimension $1$, and the decision variable $n_1=\mathcal{O}(K^2M^2)$.  For \emph{M-2''}, it has $KJ$ LMI constraints of dimension $1$, $KJ$ LMI constraints of dimension $M$, $I+L$ SOC constraint of dimension $1$, and the decision variable $n_2=\mathcal{O}(KM^3)$. As a result, for the solution with a given precision $\epsilon>0$, the worst-case complexities of Algorithm 1 and Algorithm 2 for per iteration are in the order of $\ln(1/\epsilon)\varsigma_{1,2}$, c.f. Table \ref{Comp}.

\begin{table*}
\centering
\caption{The worst-case complexities of proposed algorithms for per iteration}\label{Comp}
\begin{tabular}{|c|c|}
\hline
Algorithms & Complexity is in order of $\ln \left( 1/\varepsilon  \right)\varsigma_{\{1,2\}}$  \\ \hline
Algorithm 1 & $\varsigma_1=\sqrt{2K}\left( {{n}_{1}}K+n_{1}^{3} \right)$  \\\hline
Algorithm 2 & $\varsigma_2=\sqrt{KJ\left( M+1 \right)+2(I+L)}\cdot {{n}_{2}}\cdot \left[ KJ\left( {{M}^{3}}+{{n}_{2}}{{M}^{2}}+{{n}_{2}}+1 \right)+I+L+n_{2}^{2} \right]$  \\
\hline
\end{tabular}
\end{table*}

\section{Simulation Results}
\begin{table*}
\centering
\caption{Simulation Parameters }\label{Simulation}
\begin{tabular}{|c|c|}
\hline
Parameters & Values \\ \hline
Number of BS antennas& $N=64$  \\\hline
Sensor & $K=20,M=3$ \\\hline
Streams &$O=1,I=1,L=1,J=1$ \\\hline
Cell radius    & $500 \text{ m}$ \\\hline
RMS of reflection coefficient & $R_i=R_o=1$ \\\hline
Weighted coefficient & $\theta_{i}^{\text{sens}}=\theta_{l}^{\text{comp}}=\theta_{k,j}^{\text{sens}}=1$ \\\hline
Maximum transmit SNR at the sensors & $\mathrm{SNR}=5$ dB \\\hline
Noise powers & $\sigma_n^2=-50$ dBm \\\hline
QoS requirements & $\gamma_0=0.1$ dB, $\delta_0=0.01,\chi_0=0.01$  \\\hline
Performance priority & $\alpha_1=\alpha_2=\alpha_3=1/3$  \\\hline
\end{tabular}
\end{table*}
In this section, we conduct extensive simulations to verify the effectiveness of the proposed algorithms. Without loss of generality, all sensors are assumed being randomly distributed within the cell radius. The pass loss is modeled as $\mathrm{PL}_{\mathrm{dB}}=128.1+37.6\log_{10}(d)$ \cite{pathlossmodel}, where $d$ (km) is the distance between the transmitter and receiver. Moreover, it is assumed that all sensors have the same QoS requirements and the same maximum transmit power budget, namely $\gamma_{k,j}=\gamma_0$, $\delta_i=\delta_0$, $\chi_l=\chi_0$, and $P_{\max,k}=P_0$. We use $\mathrm{SNR}=10\log_{10}(P_0/\sigma_n^2)$ to denote the transmit signal-to-noise ratio (SNR) (in dB) at the sensors. Unless otherwise stated, the simulation parameters are summarized in Table \ref{Simulation}.

\begin{figure}[h]
 \centering
\includegraphics [width=0.45\textwidth] {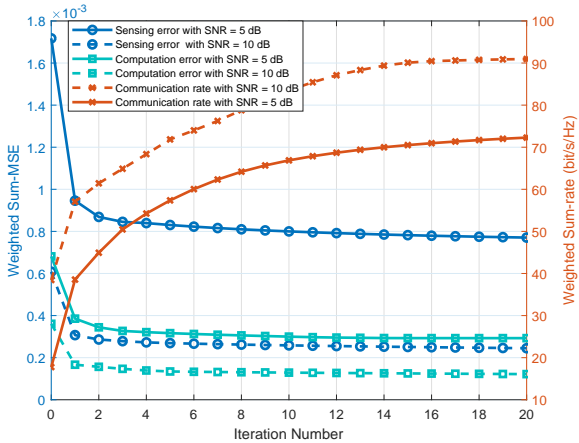}
\caption {Convergence behavior of Algorithm 1.}
\label{IterMOOP}
\end{figure}
First, we give the convergence performance of Algorithm 1 with different SNRs at the sensors. As is shown in Fig. \ref{IterMOOP}, the sensing error and the computation error decrease while the communication rate increases monotonically in the iterations, and they converge to a stable equilibrium point within few iterations on average. Hence, the implementation cost of Algorithm 1 is bearable for practical systems.

\begin{figure*}[t]
 \centering
\includegraphics [width=1\textwidth] {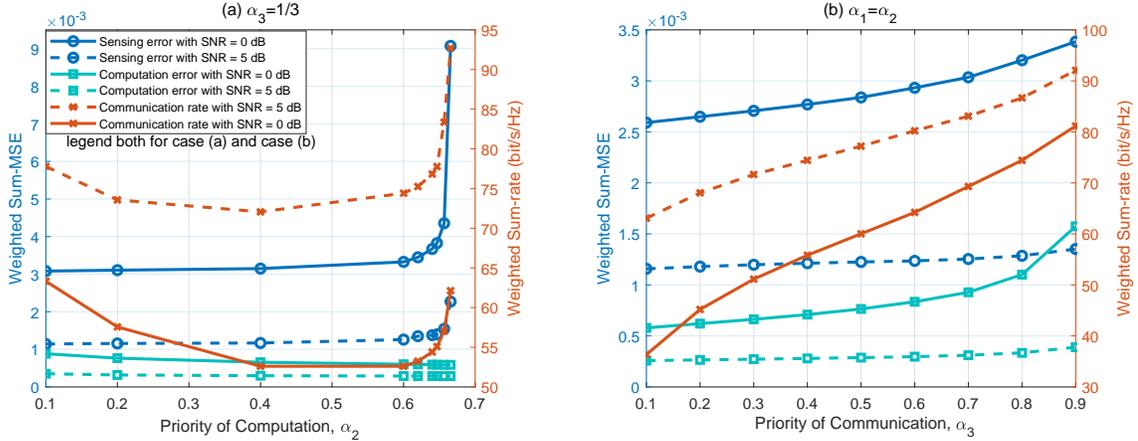}
\caption {The system performance gains versus priorities with different SNRs for Algorithm 1.}
\label{alphaMOOP}
\end{figure*}
Then, we investigate the effect of priorities of sensing $\alpha_1$, computing $\alpha_2$, and communication $\alpha_3$ of MOOP on the performance under different transmit SNRs at the sensors. Here, we present two different cases. As shown in Fig. \ref{alphaMOOP}, for case (a), we fix the weighted coefficient of communication $\alpha_3=1/3$. In this case, $\alpha_1+\alpha_2=2/3$, and then we change the value of the priority of computing $\alpha_2$ to obtain different solutions. In case (b), we fix $\alpha_1=\alpha_2$ while changing the priority of communication $\alpha_3$ to get various results.  It is observed that in case (a) with increasing of $\alpha_2$, the three performance metrics have different varying trends. Specifically, the computation error slowly decreases, the sensing error gradually increases, while the communication rate first decreases and then increases. Moreover, it is seen that in case (b) the performance of sensing and computation gradually reduces while the performance of communication gradually improves as $\alpha_3$ increases. Hence, it makes sense to select an appropriate set of priorities to balance the performance of sensing, computing, and communication. Moreover, it is found that there is a big gap between the SNR at 5 dB and 0 dB for all three performance metrics, which means increasing the SNR at the sensors can bring more gain for the overall system performance.

\begin{figure}[h]
 \centering
\includegraphics [width=0.45\textwidth] {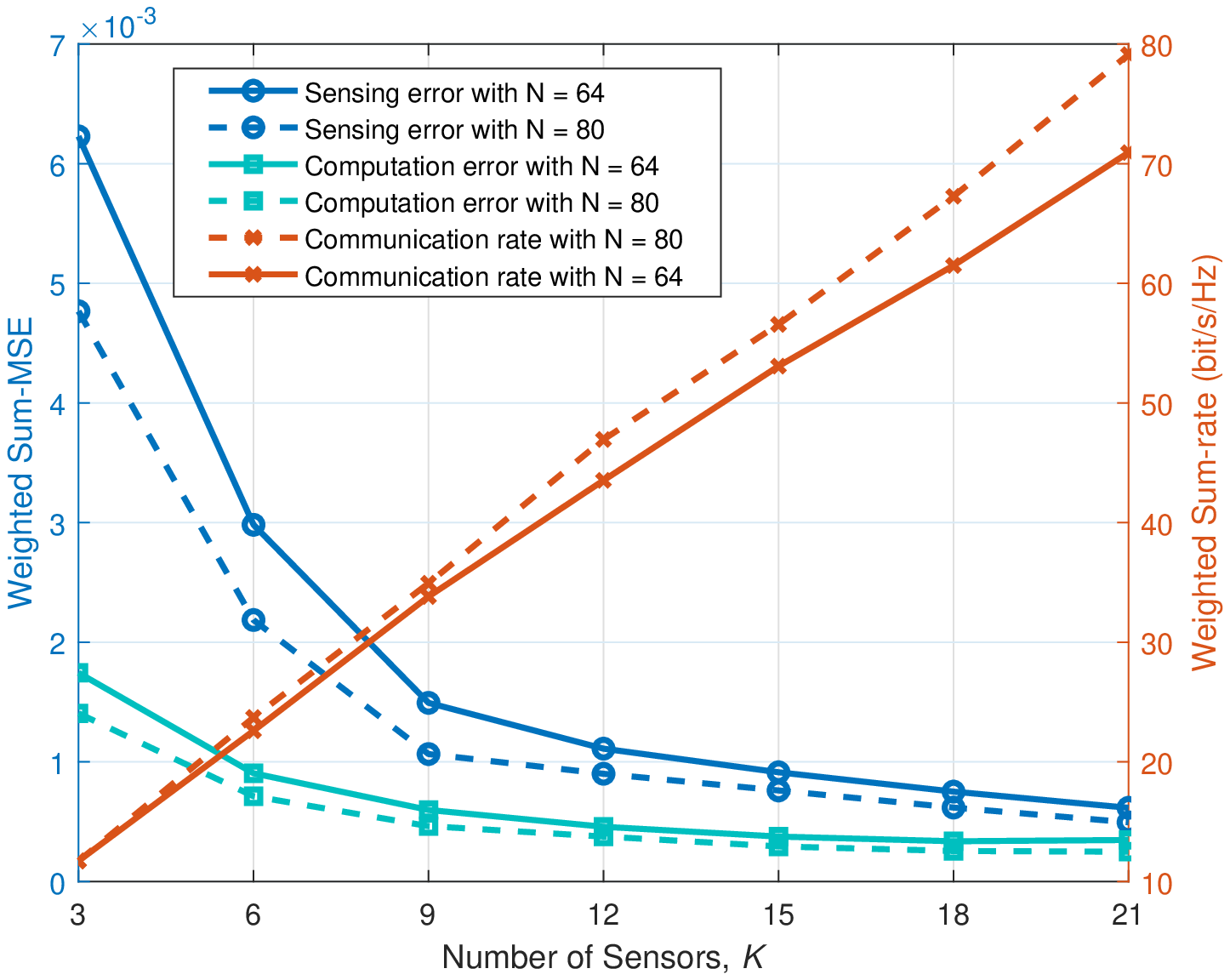}
\caption {The system performance gains versus the number of sensors with different numbers of BS antennas for Algorithm 1.}
\label{KNMOOP}
\end{figure}
Fig. \ref{KNMOOP} shows the influence of the number of sensors $K$ with different numbers of BS antennas $N$ on the performance of Algorithm 1. For a given SNR at the sensors, Algorithm 1 with more antennas at the BS can get a better overall performance. This is because a higher spatial multiplexing gain provided by the extra antennas are exploited to effectively enhance the performance. Furthermore, it is seen that with the increasing of the number of sensors, both the sensing error and the computation error diminishes while the communication rate increases. That implies the increment of the number of sensors is conducive to acquire a more accurate estimation on the reflection coefficient for the target object, a more accurate model aggregation for AirFL, and a higher weighted sum-rate for information communication, which is quite attractive for 6G wireless networks with massive sensors.

\begin{figure}[h]
 \centering
\includegraphics [width=0.45\textwidth] {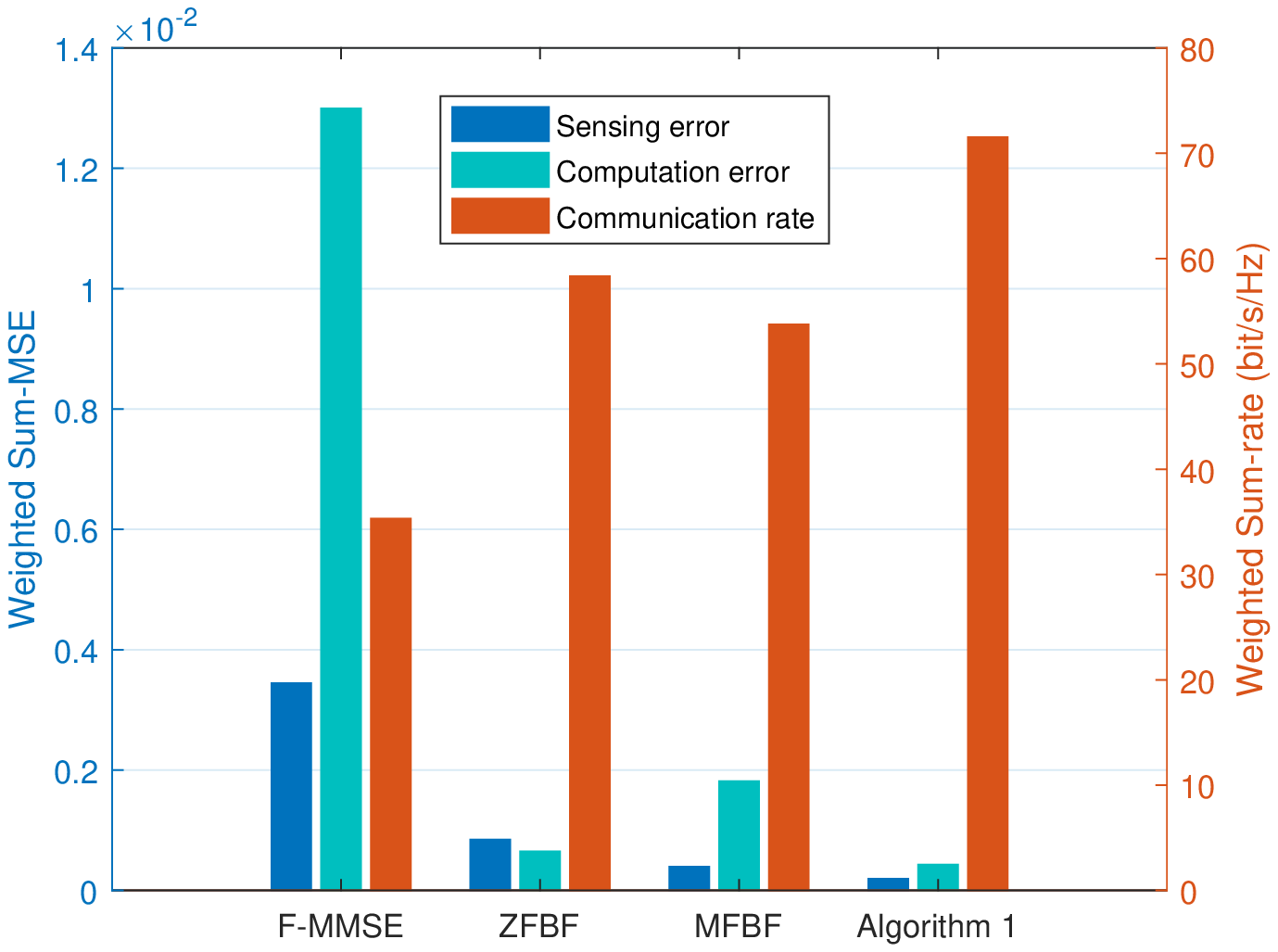}
\caption {The comparison of performance on MOOP for different algorithms.}
\label{CompareMOOP}
\end{figure}
In Fig. \ref{CompareMOOP}, we show the performance of Algorithm 1 over three baseline beamforming algorithms. They are a fixed-MMSE (F-MMSE) algorithm with fixed MMSE receivers only relevant to the channels, an AO-based match filtering beamforming (MFBF) algorithm with the match filter receivers, an AO-based zero-forcing beamforming (ZFBF) algorithm with the zero-forcing transmitters, respectively. It is seen that the F-MMSE algorithm performs the worst among the other algorithms due to the fixed receiver, which limits the performance. The MFBF algorithm outperforms the ZFBF algorithm in the performance of sensing, but performs worse in the aspects of computing and communication. Moreover, it is found that Algorithm 1 performs the best for all three performance since it obtains transmit beams and receive beams both in the optimal way, which demonstrates the superiority of Algorithm 1.

\begin{figure}[h]
 \centering
\includegraphics [width=0.45\textwidth] {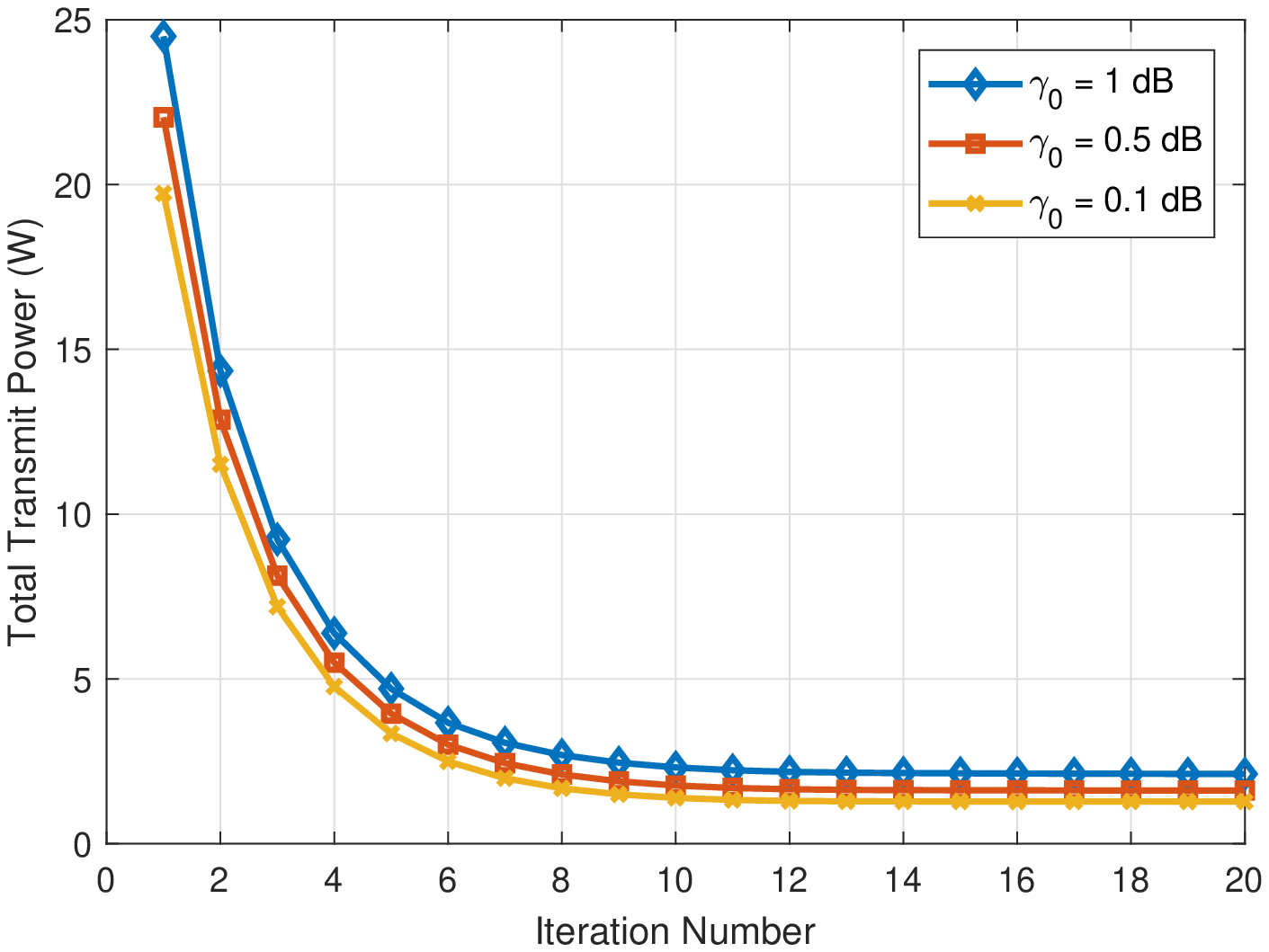}
\caption {Convergence behavior of the proposed Algorithm 2.}
\label{IterPower}
\end{figure}
Fig. \ref{IterPower} checks the convergence performance of Algorithm 2 under different required minimum SINRs of communication. It is observed that the total transmit power is progressively lowering in the iterations, and then converges within no more than 10 iterations on average under different required minimum SINRs of communication, which verifies the feasibility of Algorithm 2. %It is also demonstrated by the actual per-iteration runtime vs the number of sensors of Algorithm 2 listed in Table \ref{complexity}.

\begin{figure}[h]
 \centering
\includegraphics [width=0.45\textwidth] {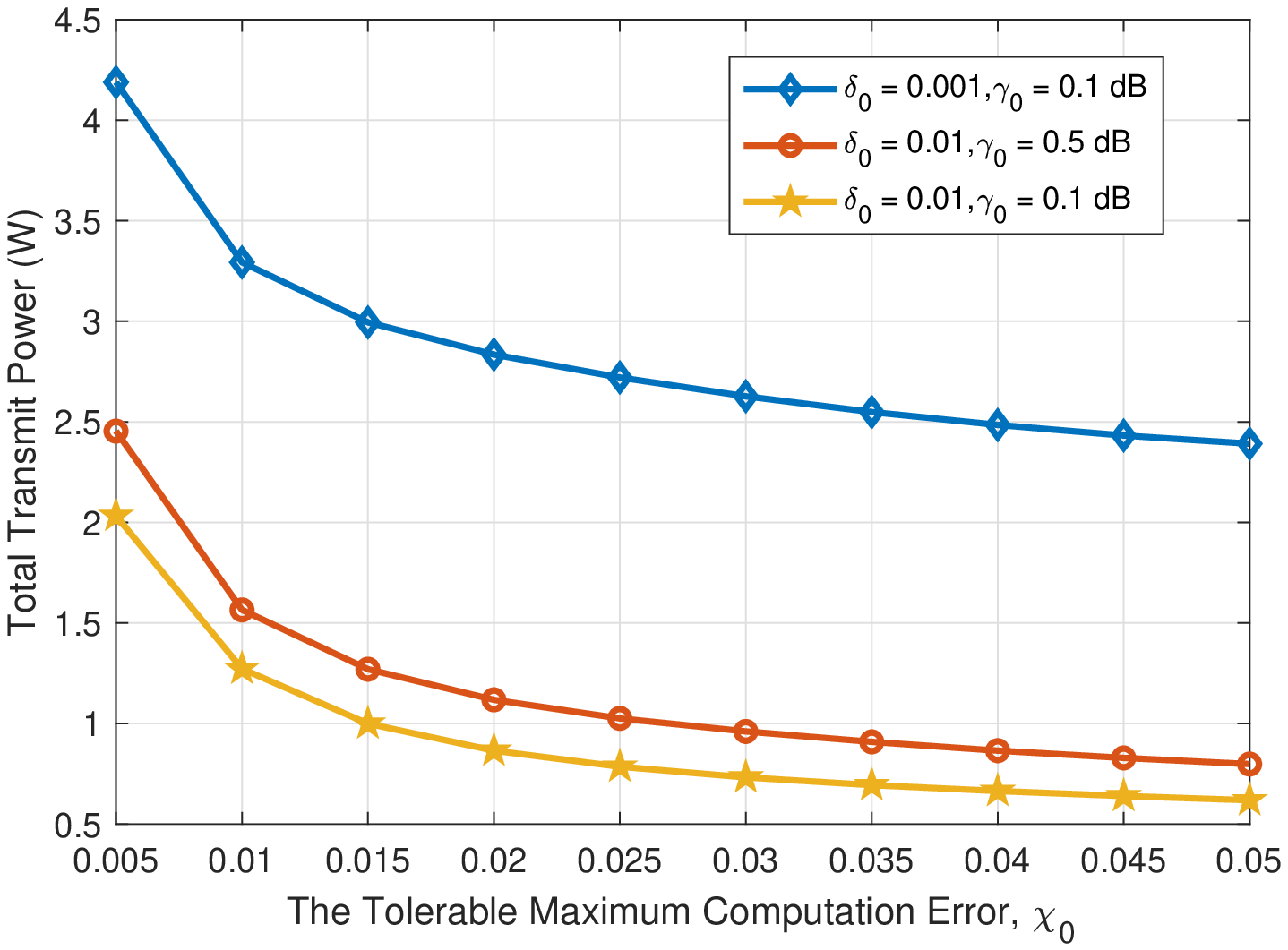}
\caption {Total transmit power versus tolerable maximum computation error with different maximum tolerable sensing error and required minimum SINR of communication for Algorithm 2.}
\label{ReqPower}
\end{figure}
Fig. \ref{ReqPower} investigates the impacts of the maximum tolerable sensing error $\delta_0$, the maximum tolerable computation error $\chi_0$, and the required minimum SINR of communication $\gamma_0$ on the total transmit power consumption. It can be seen that the total transmit power decreases as the tolerable maximum computation error increases, since a larger $\chi_0$ represents a more relax constraint on the computing accuracy resulting in less power consumption. Moreover, in the whole region of $\chi_0$, the case with $\delta_0=0.001$ and $\gamma_0=0.1$ dB requires the smallest total transmit power consumption, while the case with $\delta_0=0.01$ and $\gamma_0=0.1$ dB requires the biggest one. That means a larger $\delta_0$ is less restriction on sensing accuracy and a larger $\gamma_0$ signifies higher requirement on communication. Hence, it makes sense to loosen performance requirements appropriately for reducing the total power consumption at the sensors.

\begin{figure}[h]
 \centering
\includegraphics [width=0.45\textwidth] {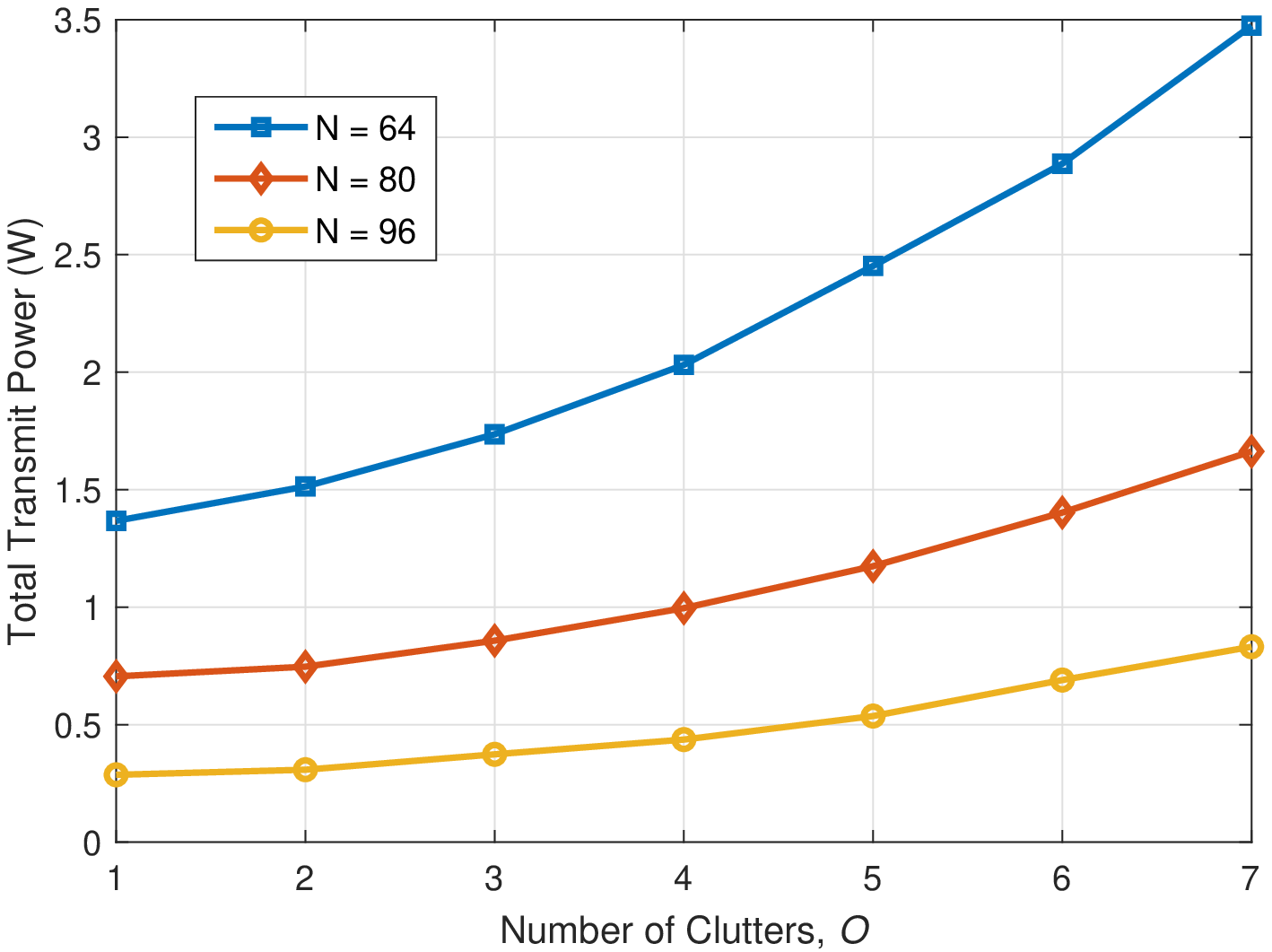}
\caption {Total transmit power versus the number of clutters with different number of BS antennas for Algorithm 2.}
\label{Clutters_N}
\end{figure}
Next, we study the influences of the numbers of clutters $O$ and BS antennas $N$ on the performance of Algorithm 2 in Fig. \ref{Clutters_N}. It is found that the total transmit power significantly increases as the number of clutters increases, since clutters as interference component is harmful to the three performance metrics resulting in more power consumption.  Moreover, Algorithm 2 with more BS antennas performs better thanks to exploiting more array gains. Besides, the performance gain gap between $N=80$ and $N=96$ is less than that between $N=64$ and $N=80$, which implies the performance enhancement provided by adding the number of BS antennas is not infinite. Notice that in practical systems, the BS equipped with more antennas could effectually decrease the total transmit power consumption, but it also leads to a higher cost on radio frequency chains. Hence, it makes sense to select a proper number of BS antennas for balancing the performance and the cost.

\begin{figure}[h]
 \centering
\includegraphics [width=0.45\textwidth] {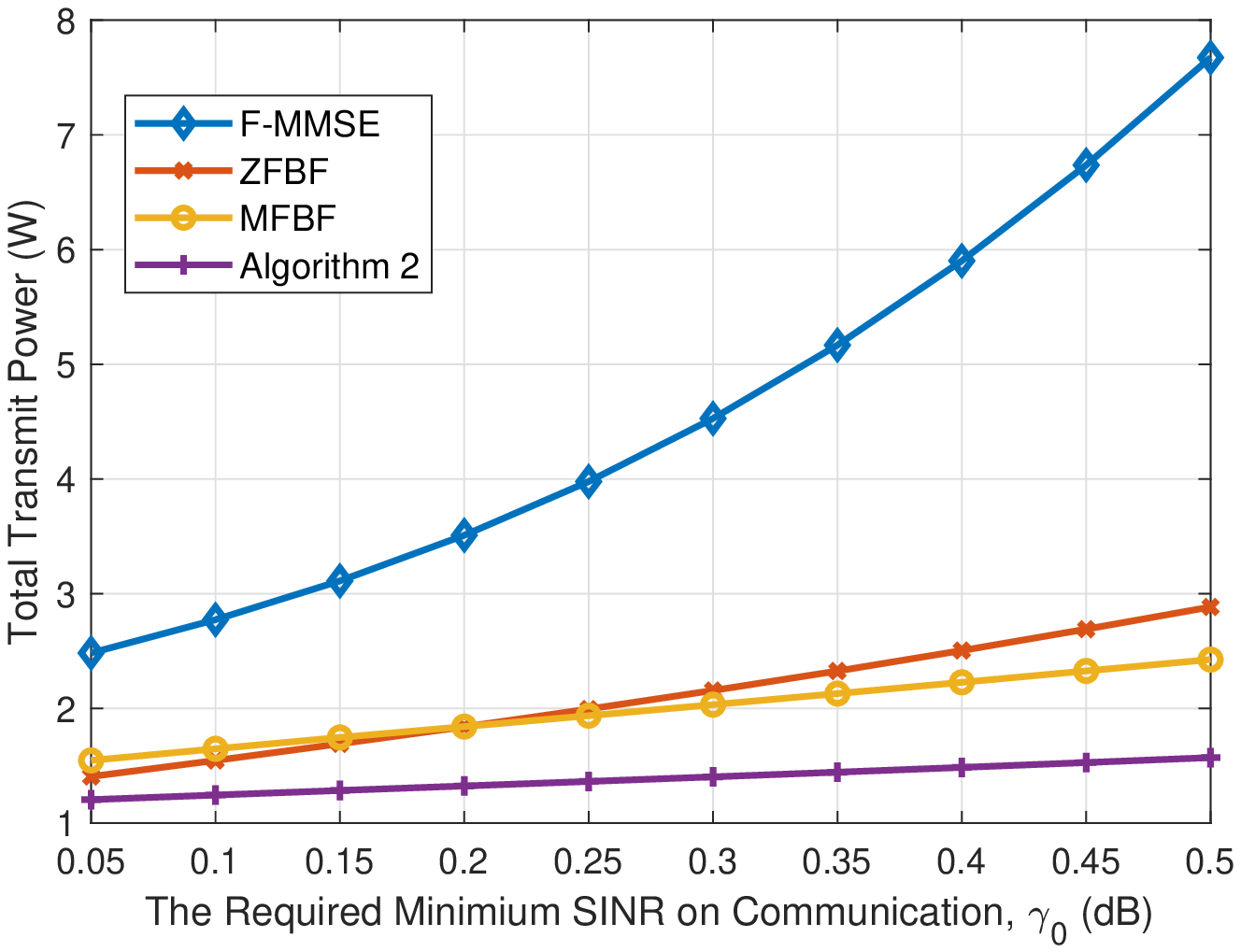}
\caption {Performance comparison of different algorithm for total transmit power minimization.}
\label{cpmparePower}
\end{figure}
Finally, we compare the performance of different algorithms in the sense of minimizing the total transmit power in Fig. \ref{cpmparePower}. There are four algorithms, namely a F-MMSE algorithm, an AO-based MFBF algorithms, an AO-based ZFBF algorithm, and Algorithm 2. For all algorithms, the total transmit power increases as the required minimum SINR on communication $\gamma_0$ increases. This is because for given sensing and computing requirements, a larger $\gamma_0$ means a more stricter requirement on communication leading to more transmit power consumption. It is evident that the F-MMSE algorithm presents the most deficient performance compared to the three AO algorithms. In addition, although the ZFBF algorithm performs better than the MFBF algorithm in the low-value region of $\gamma_0$, however, it performs worse in the high-value region of $\gamma_0$. More importantly, the proposed Algorithm 2 achieves the best performance in the whole region of $\gamma_0$, especially for high requirement of communication, which verifies the advantages of Algorithm 2 for 6G wireless networks integrating sensing, computing, and communication.

%Conclusion
\section{Conclusion}
This paper provided a comprehensive architecture for enabling 6G wireless networks to integrate sensing, computing, and communication. Two typical MOOP-based algorithms were designed from the perspectives of maximizing the weighted overall performance and minimizing the total transmit power consumption. Extensive simulations confirmed the excellent performance of the proposed algorithms. Moreover, according to the preference of applications for 6G wireless networks, it was feasible to acquire desired performance of sensing, computing, and communication by varying the system priorities. Besides, it was found that the overall performance can be improved by increasing the suitable numbers of sensors and BS antennas.

6G is still in the initial conceptual stage, and many challenges faced by ISCC has not been well addressed. In this paper, we only make a general integration of the three isolated functions of sensing, computing and communication to explore their relationship. To support future emerging various applications with ultra-high performance requirements, it is necessary to deeply integrate wireless sensing functions such as positioning, detection, and imaging with wireless transmission functions, and at the same time exploit widely distributed computing power for auxiliary processing to achieve cross-integration of sensing, computing and communication.

\begin{appendices}
\section{Proof of Theorem 1}
For the received communication signal $y_{k,j}^{\text{comm}}$ at the BS, the MSE related to the $j$-th communication signal from the $k$-th sensor is given by
\begin{eqnarray}\label{pmse1}
\!\!\!\!\!\!\!\!\!\!\!\!\!\!\!\text{MSE}_{k,j}^{\text{comm}}\!\!\!&=&\!\!\!\mathbb{E}\left\{ \left( y_{k,j}^{\text{comm}}-s_{k,j}^{\text{comm}} \right){{\left( y_{k,j}^{\text{comm}}-s_{k,j}^{\text{comm}} \right)}^{H}} \right\}\nonumber\\
  \!\!\!&+&\!\!\!\sum\limits_{i=1}^{K}{\sum\limits_{m=1}^{J}{\mathbf{u}_{k,j}^{H}{{\mathbf{H}}_{i}}{{\mathbf{c}}_{i,m}}}}\mathbf{c}_{i,m}^{H}\mathbf{H}_{i}^{H}{{\mathbf{u}}_{k,j}} \nonumber\\
  \!\!\!&+&\!\!\!\sum\limits_{i=1}^{K}{\sum\limits_{l=1}^{L}{\mathbf{u}_{k,j}^{H}{{\mathbf{H}}_{i}}{{\mathbf{b}}_{i,l}}\mathbf{b}_{i,l}^{H}\mathbf{H}_{i}^{H}{{\mathbf{u}}_{k,j}}}}\nonumber\\
  \!\!\!&+&\!\!\!\sum\limits_{i=1}^{K}{\sum\limits_{n=1}^{I}{R_{n}^{2}\mathbf{u}_{k,j}^{H}{{\mathbf{G}}_{i,n}}{{\mathbf{a}}_{i,n}}\mathbf{a}_{i,n}^{H}\mathbf{G}_{i,n}^{H}{{\mathbf{u}}_{k,j}}}} \nonumber\\
  \!\!\!&+&\!\!\!\sum\limits_{i=1}^{K}{\sum\limits_{o=1}^{O}{\sum\limits_{m=1}^{I}{R_{o}^{2}\mathbf{u}_{k,j}^{H}{{\mathbf{F}}_{i,o}}{{\mathbf{a}}_{i,m}}\mathbf{a}_{i,m}^{H}\mathbf{F}_{i,o}^{H}{{\mathbf{u}}_{k,j}}}}} \nonumber\\
  \!\!\!&+&\!\!\!\sigma_{n}^{2}\left\| {{\mathbf{u}}_{k,j}} \right\|-\mathbf{u}_{k,j}^{H}{{\mathbf{H}}_{k}}{{\mathbf{c}}_{k,j}}-\mathbf{c}_{k,j}^{H}\mathbf{H}_{k}^{H}{{\mathbf{u}}_{k,j}}+1.
\end{eqnarray}
Extracting the common factor from the above equation, we define \small{$\bm{\Xi} =\sum\limits_{i=1}^{K}{\sum\limits_{m=1}^{J}{{{\mathbf{H}}_{i}}{{\mathbf{c}}_{i,m}}}}\mathbf{c}_{i,m}^{H}\mathbf{H}_{i}^{H}+\sum\limits_{i=1}^{K}{\sum\limits_{l=1}^{L}{{{\mathbf{H}}_{i}}{{\mathbf{b}}_{i,l}}\mathbf{b}_{i,l}^{H}\mathbf{H}_{i}^{H}}}+\sum\limits_{i=1}^{K}{\sum\limits_{n=1}^{I}{R_{\text{n}}^{2}{{\mathbf{G}}_{i,n}}{{\mathbf{a}}_{i,n}}\mathbf{a}_{i,n}^{H}\mathbf{G}_{i,n}^{H}}+}\sum\limits_{i=1}^{K}{\sum\limits_{o=1}^{O}{\sum\limits_{m=1}^{I}{R_{o}^{2}{{\mathbf{F}}_{i,o}}{{\mathbf{a}}_{i,m}}\mathbf{a}_{i,m}^{H}\mathbf{F}_{i,o}^{H}}}}+\sigma _{n}^{2}\mathbf{I}$}, then (\ref{pmse1}) can be rewritten as
\begin{eqnarray}\label{pmse2}
\text{MSE}_{k,j}^{\text{comm}}&=&\mathbf{u}_{k,j}^{H}\Xi {{\mathbf{u}}_{k,j}}-\mathbf{u}_{k,j}^{H}{{\mathbf{H}}_{k}}{{\mathbf{c}}_{k,j}}-\mathbf{c}_{k,j}^{H}\mathbf{H}_{k}^{H}{{\mathbf{u}}_{k,j}}+1\nonumber\\
  \!\!\!&=&\!\!\!\left( \mathbf{u}_{k,j}^{H}-\mathbf{c}_{k,j}^{H}\mathbf{H}_{k}^{H}{{\Xi }^{-1}} \right)\Xi {{\left( \mathbf{u}_{k,j}^{H}-\mathbf{c}_{k,j}^{H}\mathbf{H}_{k}^{H}{{\Xi }^{-1}} \right)}^{H}} \nonumber\\
  \!\!\!&-&\!\!\!\mathbf{c}_{k,j}^{H}\mathbf{H}_{k}^{H}{{\Xi }^{-H}}{{\mathbf{H}}_{k}}{{\mathbf{c}}_{k,j}}+1.
\end{eqnarray}
It can be concluded that when ${{\mathbf{u}}_{k,j}}=\Xi^{-1}{{\mathbf{H}}_{k}}{{\mathbf{c}}_{k,j}}$, the $\text{MSE}_{k,j}^{\text{comm}}$ would be minimized. Consequently, the MMSE associated with the communication signal $s_{k,j}^{\text{comm}}$ is as below
\begin{eqnarray}\label{pmse3}
{{e}_{k,j}^{\text{comm}}}&=&1-\mathbf{c}_{k,j}^{H}\mathbf{H}_{k}^{H}{{\Xi }^{-H}}{{\mathbf{H}}_{k}}{{\mathbf{c}}_{k,j}}\nonumber\\
\!\!\!&=&\!\!\!\frac{{{\Xi }^{H}}-\mathbf{c}_{k,j}^{H}\mathbf{H}_{k}^{H}{{\mathbf{H}}_{k}}{{\mathbf{c}}_{k,j}}}{{{\Xi }^{H}}}\nonumber\\
\!\!\!&=&\!\!\!\frac{\mathbf{u}_{k,j}^{H}{{\Xi }^{H}}{{\mathbf{u}}_{k,j}}
-\mathbf{u}_{k,j}^{H}\mathbf{c}_{k,j}^{H}\mathbf{H}_{k}^{H}{{\mathbf{H}}_{k}}{{\mathbf{c}}_{k,j}}{{\mathbf{u}}_{k,j}}}{\mathbf{u}_{k,j}^{H}{{\Xi }^{H}}{{\mathbf{u}}_{k,j}}}\nonumber\\
\!\!\!&=&\!\!\!\frac{1}{1+{{\Gamma }_{k,j}}}.
\end{eqnarray}
It is seen that the MMSE of communication signal is the inverse of one plus SINR and thus helps to transform the objective function of \emph{S-3'}. The proof is completed.

\section{Proof of Theorem 2}
To start with, let us present two lemmas as follows:

\emph{Lemma 1}: If $\mathbf{Q}\mathbf{P}=\mathbf{0}$, based on Sylvester's rank inequality \cite{rankieq}, we have $\text{Rank}(\mathbf{Q})+\text{Rank}(\mathbf{P})\leq n$,  where $\mathbf{Q}\in {\mathbb{C}^{t\times n}}$ and $\mathbf{P}\in {\mathbb{C}^{n\times s}}, \forall t,n,s$.

\emph{Lemma 2}: For the two matrices of the same size $\mathbf{Q}$ and $\mathbf{P}$, it always holds true that $\text{Rank}(\mathbf{Q}+\mathbf{P})\leq\text{Rank}(\mathbf{Q})+\text{Rank}(\mathbf{P})$.
\begin{IEEEproof}
\begin{eqnarray}
\text{Rank}\left( \mathbf{Q}+\mathbf{P} \right)\!\!\!&=&\!\!\!\text{Rank}\left[ \begin{matrix}
   \mathbf{Q}+\mathbf{P}  \\
   \mathbf{0}  \\
\end{matrix} \right]\le \text{Rank}\left[ \begin{matrix}
   \mathbf{Q}+\mathbf{P}  \\
   \mathbf{P}  \\
\end{matrix} \right] \nonumber
=\text{Rank}\left[ \begin{matrix}
   \mathbf{Q}  \\
   \mathbf{P}  \\
\end{matrix} \right]\nonumber\\
 \!\!\!&\le&\!\!\! \text{Rank}\left( \mathbf{Q} \right)+\text{Rank}\left( \mathbf{P} \right).
\end{eqnarray}\label{prooflemma1}
\end{IEEEproof}
Then, we derive the Lagrangian function of \emph{M-2''} associated with $\mathbf{C}_{k,j}$, which is given by
\begin{eqnarray}
\mathcal{L}\left( {{\mathbf{C}}_{k,j}} \right)&=&\sum\limits_{k=1}^{K}{\left( \sum\limits_{i=1}^{I}{{{\left\| {{\mathbf{a}}_{k,i}} \right\|}^{2}}}+\sum\limits_{l=1}^{L}{{{\left\| {{\mathbf{b}}_{k,l}} \right\|}^{2}}}+\sum\limits_{j=1}^{J}{\text{tr}\left( {{\mathbf{C}}_{k,j}} \right)} \right)}\nonumber\\
&+&\sum\limits_{i=1}^{I}{{{\lambda }_{i}}}\left[ \overline{\text{MSE}}_{i}^{\text{sens}}-{{\delta }_{i}} \right]\nonumber\\
&+&\sum\limits_{l=1}^{L}{{{\beta }_{l}}}\left[ \overline{\text{MSE}}_{l}^{\text{comp}}-{{\chi }_{l}} \right]+\sum\limits_{k=1}^{K}{\sum\limits_{j=1}^{J}{{{\mu }_{k,j}} {{T }_{k,j}}}}\nonumber\\
&-&\sum\limits_{k=1}^{K}{\sum\limits_{j=1}^{J}{{{\mathbf{\Theta }}_{k,j}}{{\mathbf{C}}_{k,j}}}},
\end{eqnarray}
where ${{T}_{k,j}}=\sum\limits_{i=1,i\ne k}^{K}{\sum\limits_{n=1,n\ne j}^{J}{\text{tr}\left( \mathbf{u}_{k,j}^{H}{{\mathbf{H}}_{i}}{{\mathbf{C}}_{i,n}}\mathbf{H}_{i}^{H}{{\mathbf{u}}_{k,j}} \right)}}+{{X }_{k,j}}+\sigma _{n}^{2}{{\left\| {{\mathbf{u}}_{k,j}} \right\|}^{2}}-\frac{1}{{{\gamma }_{k,j}}}\text{tr}\left( \mathbf{u}_{k,j}^{H}{{\mathbf{H}}_{k}}{{\mathbf{C}}_{k,j}}\mathbf{H}_{k}^{H}{{\mathbf{u}}_{k,j}} \right)$. $\lambda_{i}, \beta_l, \mu_{k,j}$, and ${{\mathbf{\Theta }_{k,j}}}$ are Lagrange multipliers of constraint C5, C6, C7 and C8, respectively.  To explore the optimal solution $\mathbf{C}_{k,j}^{*}$ under the Slater's condition, we make use of the KKT conditions as follows:
\begin{subequations}
\begin{eqnarray}\label{p1}
\sum\limits_{i=1,i\ne k}^{K}{\sum\limits_{n=1,n\ne j}^{J}{\text{tr}\left( \mathbf{u}_{k,j}^{H}{{\mathbf{H}}_{i}}\mathbf{C}_{i,n}^{*}\mathbf{H}_{i}^{H}{{\mathbf{u}}_{k,j}} \right)}}+{{X}_{k,j}}+ \nonumber\\
\!\!\!\!\!\!\!\!\!\!\!\!\!\!\!\sigma _{n}^{2}{{\left\| {{\mathbf{u}}_{k,j}} \right\|}^{2}}-\frac{1}{{{\gamma }_{k,j}}}\text{tr}\left( \mathbf{u}_{k,j}^{H}{{\mathbf{H}}_{k}}\mathbf{C}_{k,j}^{*}\mathbf{H}_{k}^{H}{{\mathbf{u}}_{k,j}} \right)=0,
\end{eqnarray}
\begin{equation}\label{p2}
  \mathbf{\Theta }_{k,j}^{*}\mathbf{C}_{k,j}^{*}=\mathbf{0},
\end{equation}
\begin{eqnarray}\label{p3}
{{\nabla }_{\mathbf{C}_{k,j}^{*}}}\mathcal{L}&=&{{\mathbf{I}}_{M}}+\sum\limits_{i=1}^{I}{\lambda _{i}^{*}\mathbf{H}_{k}^{H}{{\mathbf{v}}_{i}}\mathbf{v}_{i}^{H}{{\mathbf{H}}_{k}}}+\sum\limits_{l=1}^{L}{\beta _{l}^{*}}\mathbf{H}_{k}^{H}{{\mathbf{z}}_{l}}\mathbf{z}_{l}^{H}{{\mathbf{H}}_{k}} \nonumber\\
&-&\frac{\mu _{k,j}^{*}}{{{\gamma }_{k,j}}}\mathbf{H}_{k}^{H}{{\mathbf{u}}_{k,j}}\mathbf{u}_{k,j}^{H}{{\mathbf{H}}_{k}}-\bm{\Theta }_{k,j}^{*}=\mathbf{0},
\end{eqnarray}
\begin{equation}\label{p4}
\lambda _{i}^{*}\ge 0,\beta _{l}^{*}\ge 0,\mu _{k,j}^{*}\ge 0,\bm{\Theta}_{k,j}^{*}\succeq\mathbf{0}.
\end{equation}
\end{subequations}
Since $X_{k,j}+\sigma _{n}^{2}{{\left\| \mathbf{u}_{k,j}^{H} \right\|}^{2}}>0$ in (\ref{p1}), we have ${{\mathbf{C}}_{k,j}^{*}}\neq\mathbf{0}$, which means
\begin{equation}\label{lp5.1}
  \text{Rank}(\mathbf{C}_{k,j}^{*})\ge 1.
\end{equation}
By exploiting the relationship between $\bm{\Theta}_{k,j}^{*}$ and $\mathbf{C}_{k,j}^{*}$ in (\ref{p2}), based on Lemma 1, we obtain that
\begin{equation}\label{lp5.15}
 \text{Rank}(\bm{\Theta}_{k,j}^{*})+\text{Rank}(\mathbf{C}_{k,j}^{*})\le M.
\end{equation}
Next, substituting (\ref{lp5.1}) into (\ref{lp5.15}) yields
\begin{equation}\label{lp5.2}
  \text{Rank}(\bm{\Theta} _{k,j}^{*})\le M-1.
\end{equation}
Then, based on Lemma 2, it is observed from (\ref{p3}) that
\begin{equation}\label{lp5.3}
  \text{Rank}(\bm{\Upsilon}_{k,j})+ \text{Rank}(\bm{\Theta} _{k,j}^{*})\ge \text{Rank}(\mathbf{I}_M),
\end{equation}
where \small{$\bm{\Upsilon}_{k,j}=\mathbf{H}_{k}^{H}\left( \frac{\mu _{k,j}^{*}}{{{\gamma }_{k,j}}}{{\mathbf{u}}_{k,j}}\mathbf{u}_{k,j}^{H}-\sum\limits_{i=1}^{I}{\lambda _{i}^{*}{{\mathbf{v}}_{i}}\mathbf{v}_{i}^{H}}-\sum\limits_{l=1}^{L}{\beta _{l}^{*}}{{\mathbf{z}}_{l}}\mathbf{z}_{l}^{H} \right){{\mathbf{H}}_{k}}$}. Since $\bm{\Upsilon}_{k,j} \neq \mathbf{0}$ and $\text{Rank}(\mathbf{I}_M)=M$, we obtain
\begin{equation}\label{lp5.4}
    \text{Rank}(\bm{\Theta} _{k,j}^{*})\ge M-1.
\end{equation}
Based on the relationship between (\ref{lp5.2}) and (\ref{lp5.4}), we can get $ \text{Rank}(\bm{\Theta} _{k,j}^{*})= M-1$.  Substituting it into (\ref{lp5.15}) and combining (\ref{lp5.1}), it is concluded that
\begin{equation}\label{lp5.5}
     \text{Rank}(\mathbf{C}_{k,j}^{*})=1,
\end{equation}
 which indicates the adopted SDR is tight.  The proof is completed.

\end{appendices}

%bibliography

\end{document}